\documentclass[10pt,aps,prl,twocolumn,superscriptaddress,nobibnotes,nodoi,noeprint]{revtex4-2}

\usepackage{graphicx}
\usepackage{amsmath,physics,bm,float}
\usepackage{soul}
\usepackage[colorlinks=true,citecolor=blue,urlcolor=blue,linkcolor=black]{hyperref}
\usepackage{amssymb}
\usepackage{natbib}
\usepackage{upgreek}
\usepackage{mathtools}
\usepackage{siunitx}

\usepackage[mathlines]{lineno}
\let\oldequation\equation\let\oldendequation\endequation
\renewenvironment{equation}{\linenomathNonumbers\oldequation}{\oldendequation\endlinenomath}
\let\oldalign\align\let\oldendalign\endalign
\renewenvironment{align}{\linenomathNonumbers\oldalign}{\oldendalign\endlinenomath}

\DeclareMathOperator*{\sign}{sign}
\DeclareMathOperator*{\MSD}{MSD}

\begin{document}
	
\title{Inertial active matter with Coulomb friction}

\author{Alexander P.\ Antonov}
\email{alexander.antonov@hhu.de}
\affiliation{
	Institut f{\"u}r Theoretische Physik II: Weiche Materie,
	Heinrich-Heine-Universit{\"a}t D{\"u}sseldorf, Universit{\"a}tsstra{\ss}e 1,
	D-40225 D{\"u}sseldorf, 
	Germany}

\author{Lorenzo Caprini}
\email{lorenzo.caprini@uniroma1.it}
\affiliation{
	Physics Department, University of Rome La Sapienza, Piazzale Aldo Moro 5, IT-00185 Rome, Italy}

\author{Anton Ldov}
\affiliation{
	Institut f{\"u}r Theoretische Physik II: Weiche Materie,
	Heinrich-Heine-Universit{\"a}t D{\"u}sseldorf, Universit{\"a}tsstra{\ss}e 1, 
	D-40225 D{\"u}sseldorf, 
	Germany}

\author{Christian Scholz}
\affiliation{
	Institut f{\"u}r Theoretische Physik II: Weiche Materie,
	Heinrich-Heine-Universit{\"a}t D{\"u}sseldorf, Universit{\"a}tsstra{\ss}e 1, 
	D-40225 D{\"u}sseldorf, 
	Germany}

\author{Hartmut L{\"o}wen}
\affiliation{
	Institut f{\"u}r Theoretische Physik II: Weiche Materie,
	Heinrich-Heine-Universit{\"a}t D{\"u}sseldorf, Universit{\"a}tsstra{\ss}e 1, 
	D-40225 D{\"u}sseldorf, 
	Germany}

\begin{abstract}
	Friction is central to the motion of active (self-propelled) objects such as bacteria, animals, and robots. While in a viscous fluid friction is described by Stokes's law, objects in contact with other solid bodies are often governed by more complex empirical friction laws. Here, we study active particles subject to Coulomb friction using a combination of active granular experiments and simulations, supported by theoretical predictions.
	The interplay of friction and activity forces induces a rich behavior resulting in three distinct dynamical regimes. While for low activity, Brownian motion is recovered, for large activity we observe a dynamical Stop \& Go regime that continuously switches from diffusion and accelerated motion. For greater activity, we observe a super-mobile dynamical regime characterized by a fully accelerated motion which is described by an anomalous scaling of the diffusion coefficient with the activity. These findings cannot be observed with Stokes viscous forces typical of active swimmers but are central in dry active objects.
\end{abstract}

\maketitle

\begin{figure}[htp!]
	\includegraphics[width=\linewidth]{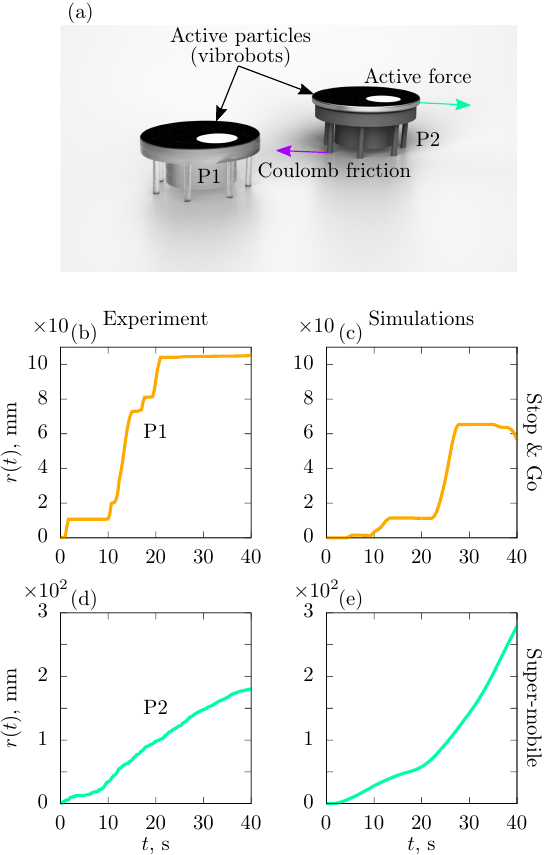}
	\caption{
		{\bf Coulomb friction-induced anomalous dynamical states}. (a)  Illustration of the experimental setup: two active vibrobots (P1 and P2, see the SM for details) governed by Coulomb friction.
		(b)-(e) Particle trajectories (radial coordinate) for different activities from experiments ((b),~(d)) and simulations ((c),~(e)). The Stop \& Go and super-mobile regimes are obtained in experiments at shaker amplitudes $A=9.381 \pm 0.046 \; \si{\mu m}$, and $A=12.825 \pm 0.089 \; \si{\mu m}$, and in simulations at activity $f_0=1, 10$. The experiments are obtained at shaker frequency $\nu=150 \; \si{Hz}$. The other simulation parameters are $\tau_0 = 100$, $\Delta_0 = 0$, $v_0 = 0.1$.
	}
	\label{fig:illustration}
\end{figure}

Friction~\cite{Bonn/etal:2014} has been studied by using empirical models starting with pioneering observations that date back to 350 A.D. (Themistius): ``It is easier to further the motion of a moving body than to move a body at rest.''
This qualitative discovery was systematically investigated by Coulomb in 1785, who noted that dry friction force depends only on the velocity direction. He developed the celebrated Coulomb's friction law~\cite{coulomb1821theorie}, which has proven to be fundamental on the macroscopic level~\cite{olsson1998friction, pennestri2016review}. At the molecular level, a further step in the study of dry friction  was taken by de Gennes and Hayakawa, who independently explored its properties for Brownian motion~\cite{deGennes:2005, hayakawa2005langevin} and suggested experimental realizations~\cite{Daniel2005}. This research has prompted further theoretical~\cite{Touchette/etal:2010, chen2014first, lequy2023stochastic, plati2023thermodynamic} and experimental investigations in the context of passive granular particles~\cite{gnoli2013granular, lemaitre2021stress}, Brownian motors~\cite{baule2012singular, chen2014large, manacorda2014coulomb,semeraro2023diffusion}, and the piston problem~\cite{sarracino2013ratchet, sano2014roles}. 


Here, we study active particles governed by Coulomb friction, whose motion is self-sustained by an activity, i.e.\ a nonequilibrium driving force with a stochastic evolution~\cite{marchetti2013hydrodynamics, Elgeti2015, bechinger2016active}. Thereby the active particles continuously inject energy into the system which is partially transformed into motion (kinetic energy) and partially irreversibly dissipated into the environment due to friction~\cite{o2022time, fodor2021irreversibility}. 
For wet systems~\cite{Zoettl/Stark:2016, lowen2020inertial}, such as active colloids and bacteria, the friction is typically linear in the velocity \cite{Romanczuk/Schimansky:2011, Romanczuk2012} due to the liquid solvent, originating from Stokes's law \cite{Elgeti_2015}. By contrast, in dry systems, such as robots and active granular particles~\cite{aranson2007swirling, kudrolli2008, kumar2014flocking, Koumakis2016, Agrawal2020, walsh2017noise, baconnier2022selective, caprini2024dynamical}, friction is generated by contact with the ground. In this case, dynamics are governed by Coulomb friction which is almost insensitive to velocity \cite{deGennes:2005, hayakawa2005langevin, Romanczuk2012}. 

Using a combination of experiments and simulations, we showcase intriguing phenomena without counterparts in systems governed by Stokes friction.
By increasing the activity, the particle switches from the standard Brownian motion dominated by white noise to a different dynamical ``Stop \& Go'' regime (Figs.~\ref{fig:illustration}~(b),~(c)), which alternates diffusive behavior to running, accelerated motion.\ Greater activity values induce an additional dynamical regime where the particle uniquely moves with an accelerated ``super-mobile'' motion  (Figs.~\ref{fig:illustration}~(d),~(e)).

An active particle with Coulomb friction is experimentally realized via a vibrobot~\cite{scholz2018inertial}, which is activated by a vertically vibrating plate (Fig.~\ref{fig:illustration}~(a)) attached to an electromagnetic shaker with frequency $\nu$ and amplitude $A$ (see the Supplemental Material (SM) for details \footnote{See Supplemental Material for details of analytical calculations, which includes Refs.~\cite{scholz2018inertial, Caroli1981, Maslov1981, Antonov2023, Landau}}). These granular particles exhibit activity due to their asymmetric design~\cite{caprini2024emergent}: the collisions between the particle legs and the plate lead to two-dimensional directed motion, with the typical speed growing as the shaker amplitude increases. Intuitively, the dynamics of this macroscopic object are governed by Coulomb friction, which is characterized by two components:
a dynamic one which decelerates an object already in motion and a static one which impedes the motion.
The dynamic part does not depend on velocity, while the static part contributes for vanishing velocity $\mathbf{v}$, which hinders the motion of a stationary object. The friction force is thus determined by two friction coefficients, which intrinsically depend on the material properties and are different for static and dynamic cases. The friction force $\boldsymbol{\sigma}(\mathbf{v})$ can be described using the two-dimensional Tustin empirical model~\cite{Lrinc/Bla:2007}:
\begin{equation}
	\label{eq:friction_force}
	\boldsymbol{\sigma}(\mathbf{v}) = \Delta_C \hat{\mathbf{v}}\left[1 + \frac{\Delta_S - \Delta_C}{\Delta_C}e^{-\frac{|\mathbf{v}|}{v_s}}\right] \,,
\end{equation}
where $\Delta_{C,S}$ are the Coulomb dynamic and static friction coefficients, respectively, with $\Delta_S \ge \Delta_C$. Here, $\hat{\mathbf{v}}=\mathbf{v}/|\mathbf{v}|$ is the normalized velocity vector, which is equal to zero when $|\mathbf{v}|=0$. The term $v_s$ is the Stribeck velocity that sets the sharpness of $\sigma(v)$ for $v\to0$~\cite{Stribeck:1902}. 
Expression~\eqref{eq:friction_force} provides a minimal description for studying dynamic and static Coulomb friction in the equation of motion. Previously, theoretical analysis focused mainly on the simpler dynamical friction case $\Delta_C = \Delta_S$, where Eq.~\eqref{eq:friction_force} reduces to $\boldsymbol{\sigma}(\mathbf{v}) = \Delta_C \hat{\mathbf{v}}$, or $\sigma(v) = \Delta_C \sign(v)$ in one dimension. 
In particular, in de Gennes's and Hayakawa's papers, emphasis was placed on velocity autocorrelation~\cite{deGennes:2005} and steady-state distribution~\cite{hayakawa2005langevin}, while Touchette \textit{et al.}~provided the exact solution~\cite{Touchette/etal:2010}. 
More recently, the role of colored noise on the velocity distribution was unveiled in the case of weak memory~\cite{Geffert/Just:2017}. Compared to~\cite{Geffert/Just:2017}, we focus on large activity and discover anomalous dynamical regimes that were not previously observed.


The dynamics of an active particle in two dimensions with mass $m$ and dry friction is minimally modeled as a Langevin equation for the particle velocity $\mathbf{v}=\dot{\mathbf{r}}$,
\begin{equation}
	m\dot{\mathbf{v}}(t) = -\boldsymbol{\sigma}(\mathbf{v}(t)) + \sqrt{2K}\boldsymbol{\xi}(t) + \mathbf{n}(t)f\,,
	\label{eq:velocity}
\end{equation}
where $\boldsymbol{\xi}(t)$ is Gaussian white noise with unit variance and $K$ determines the noise strength.
The active force is represented by the term $\mathbf{n}(t) f$, where $f$ is the activity. The term $\mathbf{n}(t)$ is an Ornstein-Uhlenbeck process with autocorrelation time $\tau$ and dynamics
\begin{equation}
	\dot{\mathbf{n}}(t) =-\frac{\mathbf{n}(t)}{\tau} + \sqrt{\frac{2}{\tau}}\boldsymbol{\eta} (t)\,,
	\label{eq:activity}
\end{equation}
where $\boldsymbol{\eta}(t)$ is Gaussian white noise with zero average and unit variance. This active force choice corresponds to the active Ornstein-Uhlenbeck particle dynamics~\cite{szamel2014self, maggi2014generalized, wittmann2017effective, caprini2018active, fily2019self, woillez2020active, martin2021statistical, PhysRevLett.129.048002}. 

\begin{figure}[t!]
	\center{\includegraphics[width=0.9\columnwidth]{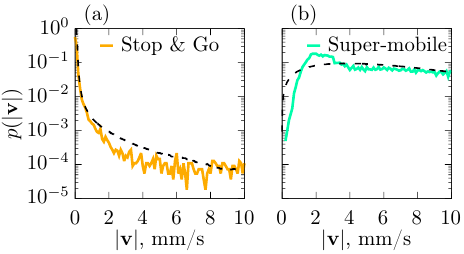}}
	\caption{\textbf{Velocity distribution $p(|\mathbf{v}|)$,} calculated from experiments (solid lines) and simulations (dashed lines). (a) The Stop \& Go regime from experiments with shaker amplitude $A=9.381 \pm 0.046 \; \si{\mu m}$ and from simulations with activity $f_0=0.3$.
		(b) The super-mobile regime observed at $A=12.825 \pm 0.089 \; \si{\mu m}$ and $f_0=4.5$. 
		In (a) and (b) the shaker frequency is $\nu=150 \; \si{Hz}$, while the other parameters are: $\tau_0 = 100, \ v_0 = 0.1, \ \Delta_0 = 0$.
	}
	\label{fig:experimental}
\end{figure}

\begin{figure*}[t!]
	\includegraphics[width=0.95\linewidth]{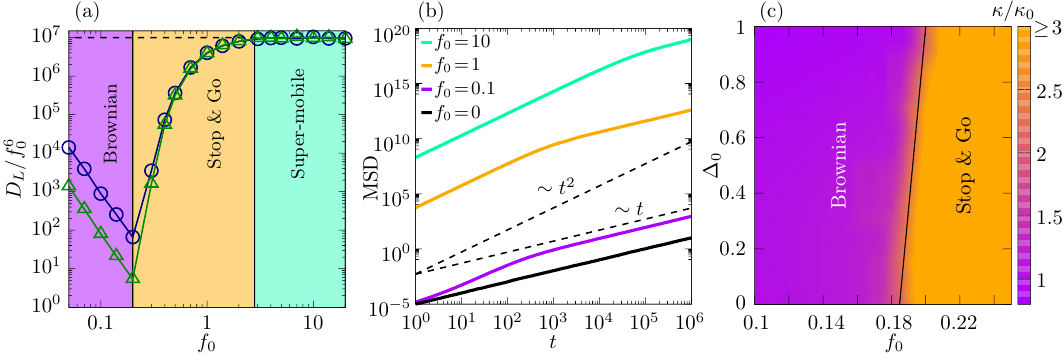}
	\caption{{\bf Stop \& Go and super-mobile states.} 
		(a) Long-time time diffusion coefficient $D_L$ rescaled by $f_0^6$ as a function of $f_0$, with $\Delta_0 = 0$ (blue circles) and $\Delta_0 = 1$ (green triangles). The parameter region corresponding to the Brownian, Stop \& Go, and super-mobile regimes are displayed in violet, yellow, and green, respectively, while the transitions are marked with vertical lines. 
		(b) Mean squared displacement $\text{MSD}= \langle ((x(t) - (x(0))^2 \rangle$, as a function of time, $t$, for different values of $f_0$ in the three dynamical states with $\Delta_0=0$. 
		Dashed black lines denote the ballistic and diffusive time scaling. 
		The colors have the same ordering as the $\MSD$ profiles.
		(c) State diagram in the plane of activity $f_0$ and relative friction coefficient $\Delta_0$. The color gradient shows the smoothed excess kurtosis, normalized with the value for the passive system. The solid black line is a guide for the eye and indicates the crossover from Brownian and Stop \& Go regimes. 
		The other parameters of the simulations are $v_0 = 0.1$ and $\tau_0 = 100$.
	}
	\label{fig:phases}
\end{figure*}

In what follows, we use $\Delta_C$, $\sqrt{\tau K}/\Delta_C$, and $\tau K/m\Delta_C$ as units of force, time, and length, respectively. 
With this choice, the system is characterized by four dimensionless parameters (see the SM for details): the reduced activity $f_0=f /\Delta_C$, which quantifies the active force effect compared to friction; the reduced noise strength $1/\tau_0=\sqrt{K/\tau}/\Delta_C$, which determines the impact of the noise kicks on the particle evolution; two friction parameters, i.e.\ the relative magnitude of the static and dynamic friction force $\Delta_0 = \Delta_S/\Delta_C - 1$ and the rescaled Stribeck velocity $v_0 = m v_s/\sqrt{\tau K}$. We set $v_0 = 0.1$ since static friction has to affect the dynamics only for small velocity, and low noise strength $1/\tau_0= 10^{-2}$, as usual in active matter experiments.

For small shaker amplitudes, i.e.,\ small activity, the active force $\mathbf{n}(t) f_0$ cannot exceed the friction force value on average, and thus, its dynamic effect is suppressed. The resulting motion is similar to the one shown by standard Brownian particles. Indeed, intuitively, the active force induces an effective dry friction coefficient which is smaller (larger) than $\Delta_C$ if the active force and the velocity have the same (opposite) sign. 
Therefore, the characteristic trajectory of this Coulomb-governed Brownian dynamical state is not qualitatively different from the usual regime displayed by a Brownian particle. 
Here, as in the passive case, static friction cannot keep the particle stationary for an arbitrary time~\cite{deGennes:2005}. In the experimental setup, the velocity in this Brownian regime is so small that it falls below the resolution limit. However, its behavior can be robustly demonstrated through simulations (see the SM).

As the shaker amplitude (and thus the activity $f_0$) is increased, the active force $\mathbf{n} (t) f_0$ is more likely to exceed the friction force value in some time interval. When this happens, the trajectory displays fast acceleration (here the particle ``Goes''). These regimes are suppressed when, by fluctuations, the active force is smaller than the Coulomb friction: when this happens, the particle behaves as a Brownian particle as observed for small $f_0$, i.e.\ the particle ``Stops''. Here, the particle is not really stuck but rather slow compared to the ``Go'' state.
This Stop \& Go behavior can be directly observed in the particle trajectory in experiments (Fig.~\ref{fig:illustration}~(b) and Supplementary Movie 1) and simulations (Fig.~\ref{fig:illustration}~(c)).
A further increase of the shaker amplitude and $f_0$ allows the active force to permanently exceed the friction value, except for the small time window when a spatial component of $\mathbf{n}(t) f_0$ reverses its direction. In this state, the friction mechanism cannot hinder the motion except for in these small time windows. This induces a different regime characterized by super-mobile behavior, where the particle continuously accelerates and suddenly decelerates. (See Fig.~\ref{fig:illustration}~(d) or Supplementary Movie 2 for experiments and Fig.~\ref{fig:illustration}~(e) for simulations).  
The difference between the dynamical regimes emerges in the steady-state distribution $p(|\mathbf{v}|)$ of the velocity modulus. Indeed, in the Stop \& Go regime, $p(|\mathbf{v}|)$ is peaked at zero (Fig.~\ref{fig:experimental}~(a)) because the particle spends a long time moving slowly. By contrast, in the super-mobile regime,  $p(|\mathbf{v}|)$ displays a bump at a large speed before slowly decaying to zero (Fig.~\ref{fig:experimental}~(b)).

In the mechanism discussed here, dimensions are not crucial (see the SM). Therefore, we systematically study the dynamics \eqref{eq:velocity}-\eqref{eq:activity} in one dimension that, in addition, we can treat analytically.
These three dynamical states can be characterized by investigating the long-time diffusion coefficient $D_L$ extracted from the mean squared displacement ($\MSD$) $\langle \left(x(t) - x(0) \right)^2 \rangle$.
Independently of the activity and Coulomb friction coefficients, this observable shows a small-time ballistic regime $\langle \left(x(t) - x(0) \right)^2 \rangle \sim t^2$ (Fig.~\ref{fig:phases}~(a)), 
which is purely induced by the activity as usual for active particles~\cite{ten2011brownian, caprini2021inertial}.
In the long-time regime, the $\MSD$ approaches a diffusive behavior $\sim t$, which is due to the random change of the active force direction present in all the dynamical states.
From here, we can extract the long-time diffusion coefficient $D_L$ that is reported as a function of $f_0$ for vanishing ($\Delta_0=0$) and non-vanishing static frictions ($\Delta_0>0$).
For small values of $f_0$ corresponding to the Brownian regime, $D_L$ scales as $f_0^2$, as expected for standard active Brownian particles (see the SM for a scaling argument).
When the Stop \& Go regime is approached, $D_L$ starts increasing faster with $f_0$. This anomalous scaling is due to the ``Go'' regimes which allows the particle to further explore the surrounding space. 
When ``Stop'' events are suppressed because of the large $f_0$ value, a scaling $D_L \sim f_0^6$ is reached in correspondence with the super-mobile state (Fig.~\ref{fig:phases}~(a)). We remark that the system switches from different dynamical states ($f_0$-scaling) via smooth crossover regimes.

The relative amplitude of static and dynamic friction $\Delta_0$ affects the diffusion properties only in the Brownian regime by decreasing $D_L$, while it leaves Stop \& Go and super-mobile states almost unchanged.
This is because static friction plays a negligible role compared to dynamical friction during a Go state since $|v|\gg v_0$, and thus $\sigma(v)\approx \Delta_C \sign(v)$.
However, we intuitively expect that an increase of the static friction, via $\Delta_0$, could affect the transition line from the Brownian to the Stop \& Go regime. 
To further characterize the role of $\Delta_0$, we focus on the steady-state velocity properties. 
Indeed, during ``Go'' states the velocity gains large values which is represented by the long tails of the velocity distribution and which we quantify by studying its fourth moment.
Specifically, we focus on the excess kurtosis, i.e.,\ the deviation of the fourth velocity moment from the Gaussian value, $\kappa = \langle v^4 \rangle/\langle v^2 \rangle^2 -3$.
We normalize this observable with the excess kurtosis $\kappa_0$ at zero activity ($f_0=0$) so that $\kappa/\kappa_0$ reads $\approx1$ in the Brownian regime and assumes values $\gg1$ in the Stop \& Go state.
Our analysis as a function of $f_0$ and $\Delta_0$ (Fig.~\ref{fig:phases}~(c)) reveals that static friction $\Delta_0>0$ hinders the transition to the Stop \& Go regime. Indeed, in this case, the active force needs to exceed a larger total dry friction to induce acceleration.

\begin{figure}[t!]	
	\centering
	\includegraphics[width=1\columnwidth]{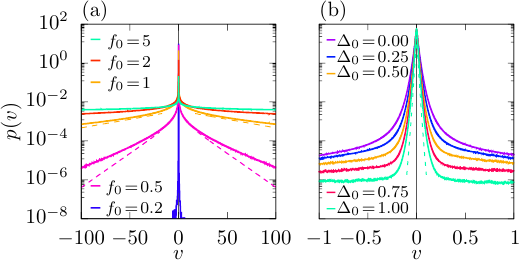}
	\caption{{\bf Probability distributions.} (a) Velocity probability distribution $p(v)$ for $\Delta_0 = 0$ for various $f_0$; (b) $p(v)$ for $f = 0.25$ for various $\Delta_0$. 
		The dashed lines in (a) correspond to the analytical prediction~\eqref{eq:active}, and dashed lines in (b) correspond to the analytical prediction~\eqref{eq:noise} up to the first order. 
		The colors have the same ordering as the density profiles.
		The other parameters of the simulations are $v_0 = 0.1$ and $\tau_0 = 100$.
	}\label{fig:4}
\end{figure}

To further shed light on the effect of activity $f_0$ and static friction $\Delta_0$, we numerically and theoretically analyze the velocity distribution $p(v)$.
As we expect from the kurtosis analysis, this observable does not show remarkable differences between the Brownian regime and the purely passive case ($f_0=0$), where $p(v)$ decays exponentially as $\propto e^{-\tau_0|v|}$. 
By resorting to path-integral techniques~\cite{Caroli1981} (see the SM for details), the expression for $p(v)$ can be generalized for small activity $f_0\ll 1$ (Brownian state) where the main contribution occurs for $|v|\ll1$ and reads
\begin{equation}    
	p(v) \propto e^{-\tau_0\left(|v|(\Delta_0 + 1) - \frac{\Delta_0}{2 v_0}v^2 - \frac{f_0^2|v|^3}{4(\Delta_0 + 1)} + \mathcal{O}(v^4)\right)} \,.
	\label{eq:noise}
\end{equation}
Here, static friction and activity have roughly the same effect, providing small deviations from the exponential tails (Figs.~\ref{fig:4}~(a)-(b)). Static friction, however, plays a pivotal role in vanishing velocities (Fig.~\ref{fig:4}~(b)).
By contrast, in the Stop \& Go state for larger values of $f_0$, the full shape of the distribution is distorted (Fig.~\ref{fig:4}~(a)). In particular, in the Stop \& Go regime, the tails slowly decay to zero, while in the super-mobile state the distribution is nearly flat with a small peak for vanishing velocity.
Remarkably, our method is able to also provide an analytical approximation for the velocity distribution $p(v)$
for large activity $f_0$, where typically $|v|\gg 1$,
\begin{subequations}
	\label{eq:active}
	\begin{equation}
		\displaystyle    p(v) \propto e^{-\frac{n_{\rm f}^2}{2}}\,.
		\label{eq:active-a}
	\end{equation}
	Here, $n_{\rm f}$ is set implicitly via the equation
	\begin{equation}
		2\tau_0\left[ n_{\rm f} f_0 - \ln(n_{\rm f} f_0) - 1\right] = v,\ n_{\rm f} > \frac{1}{f_0}\,.
		\label{eq:constraint}
	\end{equation}
\end{subequations}
Equation~\eqref{eq:active} holds in the absence of static friction $\Delta_0 = 0$, while the full expression for $\Delta_0 > 0$ is reported in the SM. However, in agreement with the previous analysis, static friction provides a negligible contribution on $p(v)$ for $|v|\gg1$. 

\begin{figure}[htp!]	
	\centering
	\includegraphics[width=\columnwidth]{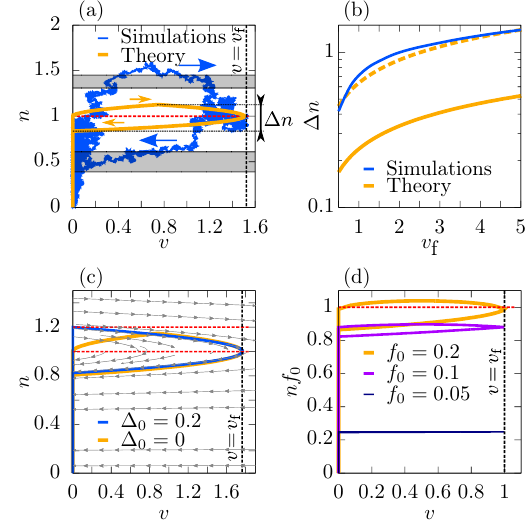}
	\caption{
		{\textbf{Analytical approach.}}
		(a),(c)-(d): Escape-and-return particle trajectory in the plane of $v$ and $n$. (a) is obtained for the maximum velocity $v_{\rm f} = 1.5$ (vertical dashed line). 
		It reports a single trajectory from numerical simulations (blue line) and from the integration of Eq.~\eqref{eq:shape} (orange line). Variances of average maximum (minimum) activity during the escape (return) $n_{\rm f}$ ($n_a$) are shown as gray rectangles.  
		(c) Cases $\Delta_0 = 0$ (without static friction) and $\Delta_0 = 0.2$ (with static friction) at $v_{\rm f} \approx 1.77$ (vertical dashed line).
		(d) Deterministic trajectories are shown for $f_0=0.05,0.1,0.2$, exploring active Brownian and Stop \& Go regimes. The curves are obtained via the numerical minimization of Eq.~\eqref{eq:traj_action} (see the SM) at $v_{\rm f} = 1$ (vertical line). 
		In (a),(c)-(d), the horizontal red lines correspond to the threshold activity levels $n= 1/(f_0 + \Delta_0)$.
		(b) Loop height $\Delta n$ for various $v_{\rm f}$, defined as the difference between maximum and minimum activity in the loop. The theoretical predictions (solid orange line) are compared with the numerical results obtained by averaging over 200 trajectories (blue). 
			By adjusting an \textit{ad hoc} multiplicative factor that is omitted in Eq.~\eqref{eq:active-a}, we show that the exponential behavior of the prediction (dashed orange line) quantitatively matches the simulations.
		(c) Analytical escape trajectories for the different static frictions $\Delta_0 = 0$ (orange line) and $\Delta_0 = 0.2$ (blue line) and the same maximum velocity $v_{\rm f} \approx 1.77$ (vertical dashed line).
		The other parameters are: $f_0=1$, $\Delta_0=0$, $v_0 = 0.1$, and $\tau_0 = 100$. 
	}\label{fig:5}
\end{figure}

Our theoretical approach allows us to predict the typical escape particle trajectory by expressing $v$ as a function of $n$ for vanishing noise (see the SM for details):
\begin{equation}
	\frac{\dd v}{\dd n} = \mp\frac{\sign(v)\tau_0}{n}\left[1 + \Delta_0 e^{-\frac{|v|}{v_0}}\right] \pm f_0 \tau_0 \,.
	\label{eq:shape}
\end{equation}
The two solutions imply that the particle displays a hysteresis-like trajectory in the $(n,v)$-plane (Fig.~\ref{fig:5}~(a)).
The particle, initially placed at $v = 0$, $n = 0$, maintains zero velocity until the active force exceeds the friction, i.e.\ at the threshold value such that $|v|=n f_0 = 1$.
From here, the particle starts accelerating (the upper sign in Eq.~\eqref{eq:shape}) and reaches a maximum velocity when the activity decreases below the threshold value. At this point, the particle slows down and relaxes to its initial state (the lower sign in Eq.~\eqref{eq:shape}).
This hysteresis-like trajectory is a visualization of the Stop \& Go regime and is purely induced by dry friction. Indeed, this behavior cannot be achieved by active dynamics governed by Stokes viscous forces for which $v \propto n$ or in the Brownian regime (small $f_0$) where the hysteresis degenerates into a back-and-forth line (Fig.~\ref{fig:5}~(d)). The hysteresis height loop, $\Delta n$ is more pronounced when higher velocities $v_{\rm f}$ are reached during the escape (Fig.~\ref{fig:5}~(b)). 
In addition, this height is increased by static friction, as shown by the escape trajectories in Fig.~\ref{fig:5}~(c). This is because the "go" state is hindered by the static friction, unlike in the case $\Delta_0=0$.

The emergence of the three dynamic states due to the interplay between activity and Coulomb friction suggests their existence in a broader range of experimental systems beyond the considered system.
Good candidates are Hexbug particles~\cite{dauchot2019dynamics, tapia2021trapped, leoni2020surfing, horvath2023bouncing, chen2023molecular} or sliding robots~\cite{Hamon/etal:2010}, where Coulomb friction can be enhanced by modifying the material properties or propulsion mechanism. Alternatively, activity can be induced on any granular object via programmed shakers that produced colored noise~\cite{kudrolli2010concentration, deseigne2010collective, soni2020phases, caprini2024emergent}. The emerging Stop \& Go and super-mobile regimes could lead to unprecedented collective phenomena for active systems ranging from giant density fluctuations to pulsating clusters. In addition, our findings can pave the way toward the development of intriguing applications in active granular matter: super-mobile active granulates could be employed for efficient spatial exploration and food search, by taking advantage of their enhanced diffusivity. Furthermore, while we have provided experimental realizations of Stop \& Go and super-mobile regimes, further miniaturization and higher resolution techniques are required to investigate the Brownian regime. A relevant example that fulfills these needs and shares similarities with the vibrobot setup is dust mitigation through vibrating surfaces~\cite{Abubakar/etal:2020}.

{\it {Acknowledgments.}}
AA, AL, CS and HL acknowledge the financial support by Deutsche
Forschungsgemeinschaft (German Research Foundation), Project LO 418/25-1.
LC acknowledges the European Union MSCA-IF fellowship for funding the project CHIAGRAM.

\clearpage
\widetext
\begin{center}
	\large{\textbf{Supplemental Material for \\ Inertial active matter with Coulomb friction}} 
	\normalsize
	
\vspace{3ex}

Alexander P.\ Antonov,$^1$ Lorenzo Caprini,$^2$ Anton Ldov,$^1$ Christian Scholz,$^1$ and Hartmut L\"owen$^1$
\small
\vspace{1ex}

$^1$\textit{Institut f{\"u}r Theoretische Physik II: Weiche Materie, \\
	Heinrich-Heine-Universit{\"a}t D{\"u}sseldorf, 
	Universit{\"a}tsstra{\ss}e 1, 
	D-40225 D{\"u}sseldorf, 
	Germany}

\vspace{.1ex}

$^2$\textit{Physics department, University of Rome La Sapienza, Piazzale Aldo Moro 5, IT-00185 Rome, Italy}
\end{center}

\setcounter{equation}{0}
\setcounter{figure}{0}
\setcounter{table}{0}
\setcounter{page}{1}
\makeatletter
\renewcommand{\theequation}{S\arabic{equation}}
\renewcommand{\thefigure}{S\arabic{figure}}
\renewcommand{\bibnumfmt}[1]{[#1]}
\renewcommand{\citenumfont}[1]{#1}

\begin{center}
\begin{minipage}{0.77\columnwidth}
	\small
	\hspace{2ex}
In this Supplemental Material, we describe the experiments reported in the main text. In addition, we provide details on the numerical study and the dimensionless parameters for the dynamics of an active particle subject to static and dynamic Coulomb friction. Finally, we derive the theoretical predictions (Eqs.~\eqref{eq:noise}-\eqref{eq:shape}) reported in the main text and discuss the method employed and the approximation involved.
\end{minipage}
\end{center}

\vspace{-2ex}
\section{Experimental setup}
\vspace{-1ex}

In the experiment, we use two active vibrobots that are manufactured out of a proprietary photopolymer using a stereolithographic 3D printer. The body of the vibrobots consists of two concentric cylinders, with a height $h_{\rm d} = 2\; \si{mm}$ and a diameter $d_{\rm d} = 15\; \si{mm}$ for the upper cylinder (vibrobot disk), and $h_{\rm c} = 4\; \si{mm}$ and $d_{\rm c} = 9\; \si{mm}$ for the lower cylinder (vibrobot core). There are seven legs suspended from the disk, inclined to the normal surface by $\theta = 4^\circ$ for the first vibrobot (P1) and by $\theta = 2^\circ$ for the second vibrobot (P2). These legs are also cylindrical in shape, with a diameter $d_l = 0.4 \; \si{mm}$ and a height $h_l = (h_0 + h_{\rm c})/\cos \theta$, where $h_0 = 1\; \si{mm}$. The total mass of P1 is $m_1 = 0.83 \pm 0.01\; \si{g}$. In comparison to P1, P2 has an additional $h_0 = 1$ \si{mm} high socket over the disk with a single steel washer attached to this socket. Except for that socket with the washer, the vibrobots P1 and P2 have the same shape (and different material). The total mass of P2 is $m_2 = 2.50 \pm 0.01 \; \si{g}$. An illustrative design of this vibrobot can be found, for example, in Ref.~\cite{scholz2018inertialS}.

The vibrobots are placed on a confined flat acrylic surface, oscillating vertically at 150 \si{Hz}. The oscillation is provided by an electromagnetic shaker, driven by an amplified signal ultimately produced with a conventional function generator. These vibrations pump energy into the inclined particle legs, resulting in the directed motion of the vibrobots when elastic energy is released. The motion generated along the vertical direction by the oscillating plate is small and affects the two-dimensional vibrobot dynamics on the plate as an effective translational noise (See Eq.~\eqref{eq:velocity} of the main text). This noise is due to the complex interplay between the vertical motion of the vibrobot due to the oscillating plate and random imperfections in the particle shape and the surface of the plate.

Each particle was marked on its top with a high-contrast label sticker. A camera displaced above the setup would capture nominally 150 images per second, ensuring a resolution of velocity comparable to the experiment's primary timescale as given through the substrate oscillation, and with a spatial resolution of 3.22 \si{px/mm}. Particle positions and orientations were later identified for each image in every recording with sub-pixel precision, using conventional image processing techniques. The error in the position and orientation detection of a stationary particle was generally found to be well below typical displacements of a moving one.

With the current experimental setup, we can reproduce Stop \& Go and super-mobile regimes, by using P1 and P2, respectively.
By contrast, we are not able to reproduce the Brownian state. Indeed, the translational noise characterizing our vibrobot is extremely small and comparable with the camera's accuracy in acquiring data. As a result, the vibrobot's motion is so minimal that it cannot be perceived by the naked eye and cannot be distinguished from the camera error in the particle position ($\sim 10^{-2}$ mm).
\subsection{Description of the movies}

\begin{itemize}
	\item 
	Supplementary Movie 1 (SupplementaryMovie1.avi) shows P1 in the Stop \& Go motion at a transitional oscillation amplitude of $9.381 \pm 0.046$ \si{\mu m}. The peak accelerations it experienced are therefore $0.849 \pm 0.004 g$, notably below what should suffice to lift it off the substrate altogether, but enough, evidently, to propel it beyond static friction.
	
	\item
	Supplementary Movie 2 (SupplementaryMovie2.avi) shows P2 in a persistent motion from an oscillation amplitude of $12.825 \pm 0.089$ \si{\mu m}, corresponding to peak accelerations of $1.161 \pm 0.008 g$. While this is nominally enough to lift the particle, the time it spends with reduced surface contact is nevertheless small (compare to the $\sim 1.7 g$ accelerations of Ref.~\cite{scholz2018inertialS}), so that the more noisy tumbling regime for which the Stokes friction active Brownian particle model applies is not assumed.
\end{itemize}

\section{Details of the numerical study}

\begin{figure}[t!]
	\center{\includegraphics[width=0.44\columnwidth]{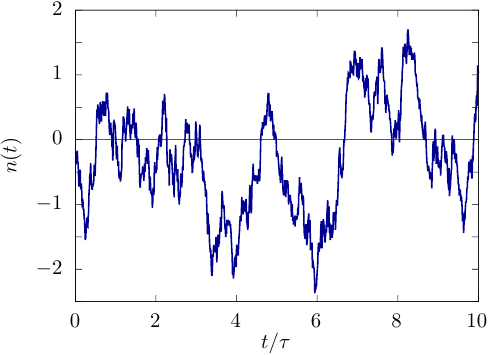}}
	\caption{Simulation of one-dimensional Ornstein-Uhlenbeck process \eqref{eq:OUP}.}
	\label{fig:OUP}
\end{figure}

In this section, we provide additional details on the numerical implementation of the equation of motion for an active particle subject to static and dynamic Coulomb friction by specifically discussing its dimensionless form.
The dynamics of an active particle with Coulomb friction (Eqs.~\eqref{eq:velocity}-\eqref{eq:activity} of the main text), is given by 
\begin{subequations}
	\begin{align}
		m\dot{\mathbf{v}}(t) &= -\boldsymbol{\sigma}(\mathbf{v}(t)) + \sqrt{2K}\boldsymbol{\xi}(t) + \mathbf{n}(t)f, \\
		\dot{\mathbf{n}}(t) & =-\frac{\mathbf{n}(t)}{\tau} + \sqrt{\frac{2}{\tau}}\boldsymbol{\eta} (t)\,\label{eq:OUP},
	\end{align}    
\end{subequations}
with 
\begin{equation}
	\boldsymbol{\sigma}(\mathbf{v}) = \Delta_C \hat{\mathbf{v}}\left[1 + \frac{\Delta_S - \Delta_C}{\Delta_C}e^{-\frac{|\mathbf{v}|}{v_s}}\right]\,.
\end{equation}
Equation~\eqref{eq:OUP} is the Ornstein-Uhlenbeck equation of motion. A typical realization of this process is shown in Fig.~\ref{fig:OUP}. The stochastic process $\mathbf{n}(t)$ is characterized by a unit steady-state variance and a typical time $\tau$, which corresponds to the autocorrelation time of the active force.

We consider $\tau K/(m\Delta_C)$ as a unit of length to rescale the particle position, $\sqrt{\tau K}/\Delta_C$ as a unit of time, and $\Delta_C$ as a unit of force, i.e.\
\begin{subequations}
	\begin{align}
		&x \rightarrow \tau K/(m\Delta_C) x,\\
		&t \rightarrow \sqrt{\tau K}/\Delta_C t,\\
		&\sigma(v) \rightarrow \Delta_C \sigma(v) \,.
	\end{align}
\end{subequations}
In this way, rescaled position, time and force are dimensionless and the corresponding dynamics read
\begin{subequations}
	\label{eq:dimensionless_dynamics2D}
	\begin{align}
		&\dot{\mathbf{v}}(t)= -\boldsymbol{\sigma}(\mathbf{v}(t)) + \sqrt{\frac{2}{\tau_0}}\boldsymbol{\xi}(t) + \mathbf{n}(t)f_0,\\
		&\dot{\mathbf{n}}(t)  =-\frac{\mathbf{n}(t)}{\tau_0} + \sqrt{\frac{2}{\tau_0}}\boldsymbol{\eta}(t),\\
		&\boldsymbol{\sigma}(\mathbf{v}) = \hat{\mathbf{v}}\left[1 + \Delta_0 e^{-\frac{|\mathbf{v}|}{v_0}}\right],
	\end{align}
\end{subequations}
or in the case of one-dimensional dynamics,
\begin{subequations}
	\label{eq:dimensionless_dynamics}
	\begin{align}
		&\dot{v}(t)= -\sigma(v(t)) + \sqrt{\frac{2}{\tau_0}}\xi(t) + n(t)f_0,\\
		&\dot{n}(t)  =-\frac{n(t)}{\tau_0} + \sqrt{\frac{2}{\tau_0}}\eta(t),\\
		&\sigma(v) = \sign(v)\left[1 + \Delta_0 e^{-\frac{|v|}{v_0}}\right],
	\end{align}
\end{subequations}

\begin{figure}[t!]
	\center{\includegraphics[width=0.9\columnwidth]{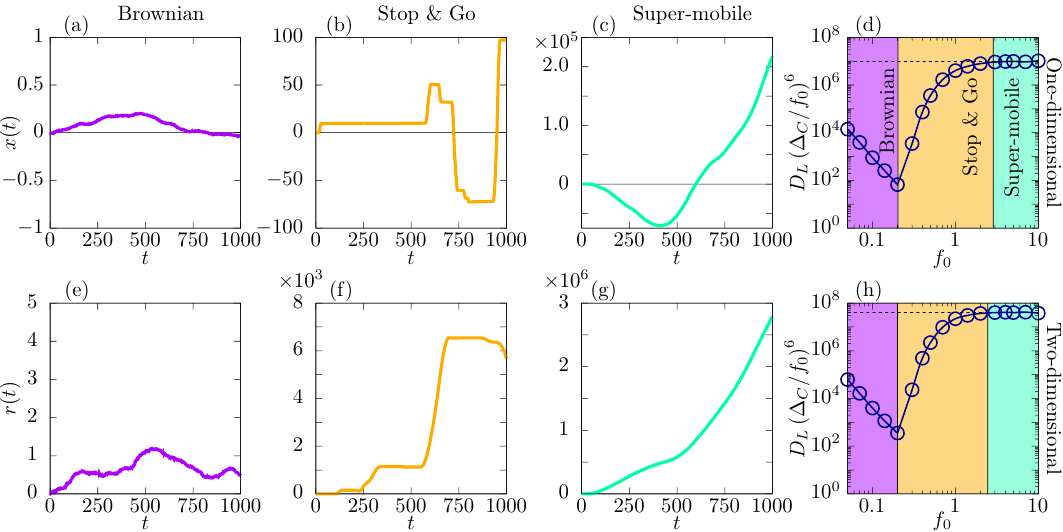}}
	\caption{(a)-(c) Simulated particle trajectories in one dimension for various activity levels: $f_0 = 0.1,1,10$, corresponding to Brownian, Stop \& Go and super-mobile regimes, respectively. (d) Long-time time diffusion coefficient $D_L$ of one-dimensional motion, rescaled by $f_0^6$ as a function of $f_0$, with $\Delta_0 = 0$. (e)-(h) Results for two-dimensional motion, with parameters matching their counterparts from panels (a)-(d).}
	\label{fig:one-two}
\end{figure}

The system is controlled by four dimensionless parameters: 
\begin{enumerate}
	\item[1)]
	the parameter $f_0 = f/\Delta_C$, i.e.\ the rescaled activity, quantifies the effect of the active force compared to friction. For $f_0\gg1$, the active force has a stronger impact compared to friction, while for $f_0\ll1$ friction dominates over the active forces. In the numerical study, $f_0$ is systematically changed, exploring both small and large activity regimes.
	\item[2)] the noise strength $1/\tau_0=\sqrt{K/\tau}/\Delta_C$, which determines the impact of the noise kick in the dynamics. This parameter is kept fixed in numerical simulations $1/\tau_0=10^{-2}$ because it is usually small in active systems. In addition, in the opposite regime, for $1/\tau_0 \gg 1$, the role of the active force is hindered. We remark that $\tau_0$ plays the role of an inverse temperature.
	\item[3)] $\Delta_0 = \Delta_S/\Delta_C - 1$ represents the relative magnitude of the static and dynamic friction force. For $\Delta_0 = 0$, the dynamics is only governed by dynamic friction, while for $\Delta_0>0$ also static friction takes place. In numerical simulations, we explore the effect of $\Delta_0$ to understand the impact of static friction of our results.
	\item[4)] the rescaled Stribeck velocity $v_0 = m v_s/\sqrt{\tau K}$
	determines the shape of the friction curve near zero velocities. The limit $v_0 \to 0$ implies that static friction is significant only at vanishing velocity when the particle does not move while increasing $v_0$ extends the relevance of the friction force to slowly moving particles. This parameter is chosen as $v_0=0.1$ in numerical simulations.
\end{enumerate}

The consideration of a one-dimensional process is motivated by the observation that the established dynamical regimes and the long-time diffusivity behavior are nearly identical between one- and two-dimensional systems, as shown in Fig.~\ref{fig:one-two}.

\section{Role of the friction in the different dynamical regimes}

In contrast to Brownian and Stop \& Go dynamical states, the active force dominates the dry friction in the super-mobile regime. However, even if small, friction still plays a significant role also in this case: indeed, its presence guarantees that the system approaches a diffusive regime.

\begin{figure}[t!]
	\center{\includegraphics[width=0.44\columnwidth]{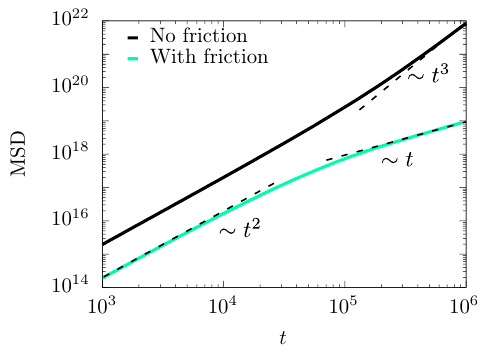}}
	\caption{Mean squared displacement (MSD) $\langle (x(t) - x(0))^2 \rangle $ as a function of time $t$ for $f = 10$ with friction ($\Delta_C = 1$, cyan) and without friction ($\Delta_C = 0$, black solid). While initially both profiles show ballistic behavior $\sim t^2$ (shown by corresponding dashed line), without friction the system goes superdiffusive ($\sim t^3$), while even small friction $\Delta_C \ll f$ is enough to establish the diffusive regime. Other simulation parameters are $\tau_0 = 100, \ v_0 = 0.1.$}
	\label{fig:MSD}
\end{figure}

As known in the literature, without the friction force, the mean squared displacement of Eq.~\eqref{eq:velocity} in the main text is characterized by a long-time superdiffusive behavior $\langle (x(t) - x(0))^2 \rangle \sim t^3$. 
This is shown in Fig.~\ref{fig:MSD} (black curve). By contrast, the presence of an arbitrarily small Coulomb friction force is enough to suppress this long-time superdiffusive behavior in favor of a diffusive regime such that $\langle (x(t) - x(0))^2 \rangle \sim t$ (cyan curve in Fig.~\ref{fig:MSD}).

Given the presence of a diffusive regime, we can estimate the long-time diffusion coefficient $D_L$ as the ratio between the square of the typical length run by the particle (persistence length) and the time that the particle needs to change direction. 
In the Brownian regime, this time coincides with the persistence time of the active force. 
Therefore, the persistence length $L_0$ in dimensionless units is given by 
\begin{equation}
	L_0 \sim v_0 \tau_0 \sim \mu_0 f_0 \tau_0 \,,
\end{equation}
where $\mu_0$ represents the bare mobility and $\tau_0$ is the persistence time. Thus, the long-time diffusion coefficient in the Brownian regime can be estimated as:
\begin{equation}
	D_L \sim \frac{L_0^2}{\tau_0} \sim \mu_0^2 f_0^2 \tau_0 \,.
\end{equation}
This argument confirms the scaling $\propto f_0^2$ numerically observed in Fig.~\ref{fig:phases}(a).

We note that the same argument can be applied also in Stop \& Go and super-mobile regimes. However, in this case, the typical time to estimate the long-time diffusion coefficient $D_L$ depends on $f_0$. Indeed, when the particle reaches extremely high velocities as in the ``Go'' states, it requires significantly more time than the persistence time to reverse its motion. 
The larger $f_0$, the larger this time.
Even if, we do not have an explanation for the scaling $D_L \sim f_0^6$, we believe that the deviation from the linear scaling ($D_L \sim f_0^2$) is reasonable as confirmed by simulations.

\section{Theoretical Prediction}

In this section, we derive the theoretical predictions, Eq.~\eqref{eq:noise} and Eqs.~\eqref{eq:active} of the main text, holding for small and large velocities, respectively.
At first, we provide a general explanation of the method which is then applied to an active particle subject to Coulomb friction, considering both the system with vanishing and non-vanishing static friction.

The system dynamics, described by the dimensionless dynamics \eqref{eq:dimensionless_dynamics}, can be expressed in a compact form by defining the vector $q(t)$ with components $(v(t), n(t))$.
In this way, for $i$ representing coordinates $v$ and $n$, we obtain
\begin{equation}
	\dot{q}_i(t) = -f_i(q(t)) + \sqrt{\frac{2}{\tau_0}}\xi_i(t)\,,
	\label{eq:dimensionless}
\end{equation}
where the general force fields $f_i$ read
\begin{subequations} 
	\label{eq:forcefield}
	\begin{align}
		f_v = & \sign(v)\left[1 + \Delta_0 e^{-\frac{|v|}{v_0}}\right] - nf_0, \\
		f_n = & \frac{n}{\tau_0}\,.
	\end{align}
\end{subequations}
To develop our theoretical argument, we consider the transition probability $\displaystyle\mathcal{P}(q_{\rm f}, t_{\rm f} | q_0, 0)$ between the initial state $q(0)=(n(0), v(0))$ at time $t=0$ and the final state $q(t_{\rm f})=(n(t_{\rm f}), v(t_{\rm f}))$ at time $t=t_{\rm f}$. 
This probability can be expressed by using the path integral formalism:
\begin{equation}
	\displaystyle\mathcal{P}(q_{\rm f}, t_{\rm f} | q_0, 0) = \int\limits_{q(0) = q_0}^{q(t_{\rm f}) = q_{\rm f}} \mathcal{D}q(t) e^{-\tau_0 S[q(t)]},
	\label{eq:path}
\end{equation}
where
\begin{equation}
	S[q(t)] = \frac{1}{4}\sum_i\int_0^{t_{\rm f}} \left( \dot{q}_i(t) + f_i(q(t)) \right)^2 \dd t + \mathcal{O}(\tau_0^{-1})
\end{equation}
is the Onsager-Machlup functional \cite{Caroli1981S}.

Let us first fix only the final velocity $v(t_{\rm f}) = v_{\rm f}$ and leave $n(t_{\rm f})$ floating. In what follows we assume that $v_{\rm f} > 0$, however, this choice is optional due to translation symmetry $v \to -v$ of Eqs.~\eqref{eq:dimensionless_dynamics}. Since the meaning of the Onsager-Machlup functional is similar to the Lagrangian of the classical system \cite{Maslov1981S}, by using analogy with the classical mechanics we can say that this functional sets the dynamical equation of motion that defines the manifold of escape paths, and the most probable escape path additionally obeys the global minimization over $n(T)$. The optimizing path that minimizes the functional satisfies the Euler-Lagrange equation,
\begin{subequations}
	\label{eq:ham}
	\begin{equation}
		\dot{q}_i(t) + f_i(q(t)) = 2 p_i(t),
		\label{eq:momentum}
	\end{equation}
	where we have introduced the analog of classical momentum as
	\begin{equation}
		\dot{p}_i(t) - p_j \partial_i f_j(q(t)) = 0 \,.
	\end{equation}
\end{subequations}
The classical analog of the Hamiltonian corresponding to Eqs.~\eqref{eq:ham} is given by
\begin{equation}
	H(p,q;t) = \sum_i \left[p_i^2(t) -f_i(q(t))p_i(t)\right]\,.
	\label{eq:hamiltonian}
\end{equation}
We remark that Eqs.~\eqref{eq:ham} can be derived by implementing the Wentzel–Kramers–Brillouin (WKB) method to the Fokker-Planck equation corresponding to the Langevin dynamics \cite{Antonov2023S}.

The Hamilton equations \eqref{eq:ham} have the first integral $\mathcal{E} = H(p,x;t)$. By combining \eqref{eq:momentum},~\eqref{eq:hamiltonian}, we can rewrite it as
\begin{equation}
	\mathcal{E} = \frac{1}{4}\sum_i \left[ \dot{q}_i^2(t) - f_i^2(q(t))\right] = \rm C \,,
	\label{eq:first_integral}
\end{equation}
where $\rm C$ is a constant term.
The probability in Eq.~\eqref{eq:path} reads
\begin{equation}
	\displaystyle\mathcal{P}(q_{\rm f}, t_{\rm f} | q_0, 0)\Big|_{ t_{\rm f} \to \infty} = P(q_{\rm f}) \propto e^{-\tau_0 S(q_{\rm f})} \,,
	\label{eq:probability}
\end{equation}
where $S$ is the analog of the classical action and can be calculated by applying the Maupertius principle \cite{LandauS},
\begin{equation}
	S = \sum_i \int\limits_{0}^{ t_{\rm f}} p_i(t) \dot{q}_i(t) \dd t - \mathcal{E}t_{\rm f} = \sum_i \int_{\rm traj} p_i \dd q_i - \mathcal{E} t_{\rm f} \,,
	\label{eq:action}
\end{equation}
where $\int_{\rm traj}$ is the integral over the trajectory of $q$.
Alternatively, if we are only interested in the shape of the trajectory, we can rewrite it as
\begin{align}
	S = &\hspace{.7ex}\mathcal{E} t_{\rm f} + \frac{1}{2}\sum_i\int \limits_{0}^{t_{\rm f}} \dd t (f_i^2(q(t)) + f_i(q(t))\dot{q}_i(t)) \nonumber \\
	= & \hspace{.7ex}\frac{1}{2}\int_{\rm traj} \left\{(2\mathcal{E} + \sum_i f_i^2(q))\sqrt{\frac{\sum_i \dd q_i \dd q_i}{4\mathcal{E} + \sum_i f_i^2(q)}} 
		+ \sum_i \left[f_i(q)) \dd q_i\right] \right\}.
	\label{eq:traj_action}
\end{align}
Since our interest is the steady-state distribution \eqref{eq:probability}, in what follows, we consider $T \to \infty$, corresponding to $\mathcal{E} = 0$ \cite{Antonov2023S}. 

\subsection{Theoretical prediction for large \textit{v}, with vanishing static dry friction}

For a particle subject to static and dry friction with the force field \eqref{eq:forcefield}, Hamilton equations correspond to four equations, for velocity $v$, stochastic process $n$, and the corresponding conjugate momenta, $p_v$ and $p_n$, respectively:
\begin{subequations}
	\begin{eqnarray}
		\dot{v} & = & \frac{2 p_v}{\tau_0} - \sign(v)\left[1 + \Delta_0 e^{-\frac{|v|}{v_0}}\right] + nf_0, \\
		\dot{n} & = & 2p_n - \frac{n}{\tau_0}, \\
		\dot{p}_v & = & p_v \left[2\delta(v)(1+\Delta_0) - \frac{\Delta_0}{v_0} e^{-\frac{|v|}{v_0}}\right], \\
		\dot{p}_n & = & \frac{p_n}{\tau_0} - p_v f_0 \,,
		\label{eq:hamilton}
	\end{eqnarray}
\end{subequations}
where $\delta(v)$ is a Dirac delta function obtained from the derivative of $\sign(v)$.
Those dynamical equations are not solvable in quadratures due to the discontinuity in the friction force profile so that this global minimization can be performed only numerically. 
We therefore reformulate the question about the velocity distributions: which factor, noise or activity, exerts greater impact when the object acquires velocity? 
In the considered case of low noise the answer is that for attaining small velocities near zero, noise plays the predominant role, similar to the passive case where noise is the sole mechanism initiating movement. For the larger velocities, this is no longer the case since activity emerges as the primary contributing factor. With this understanding, we can approximate global minimization through constrained minimization. The constraint is that the action gained along one of the axes is negligible, depending on which mechanism dominates in velocity gaining. For $v_{\rm f} \gg 1$ the dominant mechanism is activity, so the action is zero along $v$-axis. Using the corresponding ansatz $p_v = 0$, we obtain two solutions of Eqs.~\eqref{eq:hamilton}, one of which corresponds to the escape part of the trajectory path,
\begin{subequations}
	\begin{align}
		\dot{v} & = -\sign(v)\left[1 + \Delta_0 e^{-\frac{|v|}{v_0}}\right] + nf_0, \label{eq:ea} \\
		\dot{n} & = p_n = \frac{n}{\tau_0}, \label{eq:eb} \\
		p_v & = 0 \,,
	\end{align}
	\label{eq:escape}
\end{subequations}
and another corresponding to the relaxation part (gradient-path):
\begin{subequations}
	\begin{align}
		\dot{v} & = -\sign(v)\left[1 + \Delta_0 e^{-\frac{|v|}{v_0}}\right] + nf_0, \label{eq:ra}\\
		\dot{n} & = -\frac{n}{\tau_0}, \label{eq:rb}\\
		p_v & = p_n = 0 \,.
	\end{align}
	\label{eq:return}
\end{subequations}
We note that the return part is always given by $p_n = p_v = 0$, as this is the solution for the generic form of Hamilton equations \eqref{eq:ham}. At the same moment, this return part also does not contribute to probability Eqs.~\eqref{eq:probability}, as the corresponding contribution in \eqref{eq:action} is equal to zero \cite{Antonov2023S}.

Since we are interested only in the trajectory shape, we rewrite Eqs.~\eqref{eq:escape},~\eqref{eq:return} as
\begin{equation}
	\frac{\dd v}{\dd n} = \mp\frac{\sign(v)\tau_0}{n}\left[1 + \Delta_0 e^{-\frac{|v|}{v_0}}\right] \pm f_0 \tau_0,
	\label{eq:shapy}
\end{equation}
with upper sign corresponding to escape part \eqref{eq:escape}, and lower sign corresponding to return part \eqref{eq:return}.

Let us first consider $\Delta_0 = 0$. The trajectory consists of 4 parts. Initially, in the first part, activity increases according to Eq.~\eqref{eq:eb}, while velocity remains constant at zero. Subsequently, during the second part, velocity begins to increase when $n > 1/f_0$, driven by the positive acceleration $\dot{v}$ in Eq.~\eqref{eq:ea}. The switch to the third part occurs when the escape trajectory switches to the return trajectory at $n = n_{\rm f}$, where $n_{\rm f}$ is implicitly determined by the maximum velocity $v_{\rm f}$ attained. During this part, activity starts to decline according to Eq.~\eqref{eq:rb}, yet the particle continues to accelerate until the point when $\dot{v} = 0$ as described in Eq.~\eqref{eq:ra} for $n = 1/f_0$. Following this, the particle decelerates, eventually reaching zero velocity at $n = n_{\rm a}$. Following that, in the fourth part, velocity remains at 0 while activity relaxes to 0 as well. In summary:
\begin{eqnarray}
	v(n)= \begin{cases} 0, \ n \in [0;\frac{1}{f_0}[, \\ 
		\tau_0\left[nf_0 - \ln(nf_0) - 1 \right], \ n \in [\frac{1}{f_0};n_{\rm f}[, \\
		v_{\rm f} - \tau_0\left[nf_0 - \ln(nf_0) - 1 \right], \ n \in [n_{\rm f};n_{\rm a}[, \\
		0, \ n \in [n_{\rm a};0[.
	\end{cases}
\end{eqnarray}
with 
\begin{eqnarray}
	2\tau_0\left[ n_{\rm f} f_0 - \ln(n_{\rm f} f_0) - 1\right] = v_{\rm f},\ n_{\rm f} > \frac{1}{f_0}, \label{eq:vf1}\\
	\tau_0\left[ n_{\rm a} f_0 - \ln(n_{\rm a} f_0) - 1\right] = v_{\rm f},\ n_{\rm a} < \frac{1}{f_0}\,.
\end{eqnarray}
Combining \eqref{eq:probability},~\eqref{eq:action},~\eqref{eq:escape},~\eqref{eq:return}, we obtain
\begin{equation}
	p(v_{\rm f}) \propto e^{-\frac{n_{\rm f}^2(v_{\rm f})}{2}} \,.
	\label{eq:prob}
\end{equation}
with $n_{\rm f}$ depending on $v_{\rm f}$ implicitly via Eq.~\eqref{eq:vf1}. Equation~\eqref{eq:prob} corresponds to Eqs.~\eqref{eq:active} of the main text.

\subsection{Theoretical prediction for large \textit{v}, with static dry friction}
Solving Eq.~\eqref{eq:shapy} for $\Delta_0 \ne 0$ is less trivial. Due to the nature of the static friction, here motion starts for activity $n=1/f_0(1 + \Delta_0)$. Since then friction decreases for higher velocity, this activity is \textit{a priori} enough to reach a certain velocity (which can be also larger than $v_{\rm f}$), even if the escape part switches to return immediately after the initiation of motion. The activity corresponding to the maximum velocity $v_{\rm max}$ is obtained from Eq.~\eqref{eq:ra} for $\dot{v} = 0$. Solving Eq.~\eqref{eq:shapy} before these conditions, we obtain
\begin{equation}
	v(n) = 
	\begin{cases}
		0, & \text{if } n \in \left[0, \frac{1 + \Delta_0}{f_0}\right[, \vspace{.4ex} \\ \vspace{.4ex}
		v_e(n), & \text{if } n \in \left[\frac{1 + \Delta_0}{f_0}, n_{\rm f}\right[, \ v_e\left( \frac{1 + \Delta_0}{f_0}\right) = 0, \\
		v_r(n), & \text{if } n \in [n_{\rm f}, n_{\rm a}[, \ v_r(n_{\rm a}) = 0, \ v_r\left( \frac{1 + \Delta_0 e^{-\frac{v_{\rm f}}{v_0}}}{f_0}\right) = v_{\rm max}, \\
		0, & \text{if } n \in [n_{\rm a}, 0[.
	\end{cases}
	\label{eq:fear}
\end{equation}
where escape and return parts $v_{e,r}(n)$ read
\begin{align*}
	v_e(n) &= n f_0 \tau_0 + v_0 \ln \Bigg[ -\frac{\Delta_0 \tau_0 E_{1-\frac{\tau_0}{v_0}}\left(\frac{n f_0 \tau_0}{v_0}\right)}{v_0} + n^{-\frac{\tau_0}{v_0}} \left( \exp\left[ -\frac{\tau_0 \left\{1+\Delta_0 + \ln\left(\frac{f_0}{1+\Delta_0}\right)\right\}}{v_0} \right] \right. \\
	& + \left. \frac{1}{f_0} \left\{ \frac{v_0 f_0}{\Delta_0 \tau_0} \right\}^{1 - \frac{\tau_0}{v_0}} \Gamma \left( \frac{\tau_0}{v_0}, \frac{\tau_0[1 + \Delta_0]}{v_0} \right) \right) \Bigg], \\
	v_r(n) &= -n f_0 \tau_0 + v_0 \ln \Bigg[ -\frac{\Delta_0 \tau_0 E_{1+\frac{\tau_0}{v_0}}\left(-\frac{n f_0 \tau_0}{v_0}\right)}{v_0} \\
	& + n^{\frac{\tau_0}{v_0}} \left( e^{\frac{v_{\rm max} + \tau_0 - \Delta_0 \tau_0 e^{-\frac{v_{\rm max}}{v_0}}}{v_0} } \left(-\frac{1-\Delta_0 e^{-\frac{v_{\rm max}}{v_0}}}{f_0}\right)^{-\frac{\tau_0}{v_0}} \right. - \left. \frac{1}{f_0} \left\{ -\frac{\Delta_0 \tau_0 f_0}{v_0} \right\}^{1 + \frac{\tau_0}{v_0}} \Gamma \left(- \frac{\tau_0}{v_0}, \frac{\tau_0[1 + e^{-\frac{v_{\rm max}}{v_0}} \Delta_0]}{v_0} \right) \right) \Bigg], \\
	v_e(n_{\rm f}) &= v_r(n_{\rm f}) \,.
\end{align*}
Here, $E_n(x) = x^n \Gamma(1-n,x)$ is the exponential integral, with $\Gamma(a,x)=\int_x^{\infty} t^{a-1} e^{-t} \dd t$ being the incomplete Gamma-function.

As we have mentioned before, for static friction during the initiation of motion, the particle gains (at least) some velocity $v_{\rm min}$ in any case. It is derived from the condition of instantaneous transition from escape to return at the moment of initiation and is defined by the formula:
\begin{equation}
	v_{\rm min} = -f_0 \tau_0 + \ v_0 \ln \left[ e^{\frac{\tau_0 \left(f_0 + \Delta_0 f_0 - \ln(1 + \Delta_0) \right)}{v_0} } + \frac{\Delta_0 \tau_0 \left(E_{1+\frac{\tau_0}{v_0}}\left(-\frac{f_0 \tau_0}{v_0}\right)\right) - \left\{-\frac{\tau_0 f_0}{v_0}\right\}^{\frac{\tau_0}{v_0}}\Gamma\left(- \frac{\tau_0}{v_0}, \frac{-f_0 \tau_0(1 + \Delta_0)}{v_0}\right)}{v_0}\right].
\end{equation}
Thus, the maximum velocity gained during the escape is the greater of $v_{\rm min}$ and the pre-set $v_{\rm f}$:
\begin{equation}
	v_{\rm max} = \max \left\{ v_{\rm min}, v_{\rm f} \right\}, \\
\end{equation}
The probability distribution is obtained identically to Eq.~\eqref{eq:prob}, with $n_{\rm f}$ now depending implicitly on $v_{\rm f}$ via Eq.~\eqref{eq:fear}.

\subsection{Theoretical prediction for small \textit{v}}

For noise-induced velocities $v_{\rm f} \ll 1$, the analogous ansatz $p_n = 0$ is not helpful for the solution for the escape part. Therefore, we make another assumption using a similar idea: we assume that $n_{\rm f}$ in the final state ${\bf{q}}_f$ which minimizes the action is small, $n_{\rm f} \ll 1$, and that the trajectory connecting initial and final state is a line. As a consequence, this trajectory can be parameterized as $n = \alpha v$. The action is then minimized if the angle $\alpha_0$ fulfills the following minimization condition:
\begin{equation}
	\frac{\partial S}{\partial \alpha}\Bigg|_{\alpha = \alpha_0} = 0,
\end{equation}
so that Eq.~\eqref{eq:traj_action} then reads:
\begin{equation}
	S = \frac{1}{2}\int\limits_0^{v_{\rm f}} \left\{ \sqrt{\left[(1 + \Delta_0 e^{-\frac{v}{v_0}}-\alpha v f_0)^2 + \frac{\alpha^2v^2}{\tau_0^2}\right](\alpha^2+1)} + (1 - \alpha v f_0) + \frac{\alpha^2 v}{\tau_0}\right\} \dd v \,.
\end{equation}
As this integral cannot be integrated analytically, we calculate it as Maclaurin series for small $n_{\rm f}, v_{\rm f}$. In this limit, the angle $\alpha_0$ minimizing this action is
\begin{equation}
	\alpha_0 = \frac{v_{\rm f} f_0}{1 + \Delta_0} + \mathcal{O}(v_{\rm f}^2) \,.
\end{equation}
By calculating action for this $\alpha_0$ and substituting it to Eq.~\eqref{eq:probability}, we obtain an expression for the velocity distribution $p(v)$ holding for $|v|\ll1$:
\begin{equation}
	\displaystyle	p(v) \propto e^{-\tau_0\left(|v|(\Delta_0 + 1) - \frac{\Delta_0}{2 v_0}v^2 - \frac{f_0^2|v|^3}{4(\Delta_0 + 1)} + \mathcal{O}(v^4)\right)} \,.
	\label{eq:noisy}
\end{equation}
Equation~\eqref{eq:noisy} corresponds to the prediction~\eqref{eq:noise} of the main text.
In agreement with our intuition, we additionally observe that for small $v$, the expression for $p(v)$ differs from that of a passive particle subject to dry friction only for subdominant, third-order terms in power of $v$.


\begin{thebibliography}{67}%
	\makeatletter
	\providecommand \@ifxundefined [1]{%
		\@ifx{#1\undefined}
	}%
	\providecommand \@ifnum [1]{%
		\ifnum #1\expandafter \@firstoftwo
		\else \expandafter \@secondoftwo
		\fi
	}%
	\providecommand \@ifx [1]{%
		\ifx #1\expandafter \@firstoftwo
		\else \expandafter \@secondoftwo
		\fi
	}%
	\providecommand \natexlab [1]{#1}%
	\providecommand \enquote  [1]{``#1''}%
	\providecommand \bibnamefont  [1]{#1}%
	\providecommand \bibfnamefont [1]{#1}%
	\providecommand \citenamefont [1]{#1}%
	\providecommand \href@noop [0]{\@secondoftwo}%
	\providecommand \href [0]{\begingroup \@sanitize@url \@href}%
	\providecommand \@href[1]{\@@startlink{#1}\@@href}%
	\providecommand \@@href[1]{\endgroup#1\@@endlink}%
	\providecommand \@sanitize@url [0]{\catcode `\\12\catcode `\$12\catcode
		`\&12\catcode `\#12\catcode `\^12\catcode `\_12\catcode `\%12\relax}%
	\providecommand \@@startlink[1]{}%
	\providecommand \@@endlink[0]{}%
	\providecommand \url  [0]{\begingroup\@sanitize@url \@url }%
	\providecommand \@url [1]{\endgroup\@href {#1}{\urlprefix }}%
	\providecommand \urlprefix  [0]{URL }%
	\providecommand \Eprint [0]{\href }%
	\providecommand \doibase [0]{https://doi.org/}%
	\providecommand \selectlanguage [0]{\@gobble}%
	\providecommand \bibinfo  [0]{\@secondoftwo}%
	\providecommand \bibfield  [0]{\@secondoftwo}%
	\providecommand \translation [1]{[#1]}%
	\providecommand \BibitemOpen [0]{}%
	\providecommand \bibitemStop [0]{}%
	\providecommand \bibitemNoStop [0]{.\EOS\space}%
	\providecommand \EOS [0]{\spacefactor3000\relax}%
	\providecommand \BibitemShut  [1]{\csname bibitem#1\endcsname}%
	\let\auto@bib@innerbib\@empty
	\bibitem [{\citenamefont {Fall}\ \emph {et~al.}(2014)\citenamefont {Fall},
		\citenamefont {Weber}, \citenamefont {Pakpour}, \citenamefont {Lenoir},
		\citenamefont {Shahidzadeh}, \citenamefont {Fiscina}, \citenamefont
		{Wagner},\ and\ \citenamefont {Bonn}}]{Bonn/etal:2014}%
	\BibitemOpen
	\bibfield  {author} {\bibinfo {author} {\bibfnamefont {A.}~\bibnamefont
			{Fall}}, \bibinfo {author} {\bibfnamefont {B.}~\bibnamefont {Weber}},
		\bibinfo {author} {\bibfnamefont {M.}~\bibnamefont {Pakpour}}, \bibinfo
		{author} {\bibfnamefont {N.}~\bibnamefont {Lenoir}}, \bibinfo {author}
		{\bibfnamefont {N.}~\bibnamefont {Shahidzadeh}}, \bibinfo {author}
		{\bibfnamefont {J.}~\bibnamefont {Fiscina}}, \bibinfo {author} {\bibfnamefont
			{C.}~\bibnamefont {Wagner}},\ and\ \bibinfo {author} {\bibfnamefont
			{D.}~\bibnamefont {Bonn}},\ }\href
	{https://doi.org/10.1103/PhysRevLett.112.175502} {\bibfield  {journal}
		{\bibinfo  {journal} {Phys. Rev. Lett.}\ }\textbf {\bibinfo {volume} {112}},\
		\bibinfo {pages} {175502} (\bibinfo {year} {2014})}\BibitemShut {NoStop}%
	\bibitem [{\citenamefont {Coulomb}(1821)}]{coulomb1821theorie}%
	\BibitemOpen
	\bibfield  {author} {\bibinfo {author} {\bibfnamefont {C.~A.}\ \bibnamefont
			{Coulomb}},\ }\href@noop {} {\emph {\bibinfo {title} {Th{\'e}orie des
				machines simples en ayant {\'e}gard au frottement de leurs parties et {\`a}
				la roideur des cordages}}}\ (\bibinfo  {publisher} {Bachelier, Paris},\ \bibinfo
	{year} {1821})\BibitemShut {NoStop}%
	\bibitem [{\citenamefont {Olsson}\ \emph {et~al.}(1998)\citenamefont {Olsson},
		\citenamefont {{\AA}str{\"o}m}, \citenamefont {De~Wit}, \citenamefont
		{G{\"a}fvert},\ and\ \citenamefont {Lischinsky}}]{olsson1998friction}%
	\BibitemOpen
	\bibfield  {author} {\bibinfo {author} {\bibfnamefont {H.}~\bibnamefont
			{Olsson}}, \bibinfo {author} {\bibfnamefont {K.~J.}\ \bibnamefont
			{{\AA}str{\"o}m}}, \bibinfo {author} {\bibfnamefont {C.~C.}\ \bibnamefont
			{De~Wit}}, \bibinfo {author} {\bibfnamefont {M.}~\bibnamefont
			{G{\"a}fvert}},\ and\ \bibinfo {author} {\bibfnamefont {P.}~\bibnamefont
			{Lischinsky}},\ }\href {https://doi.org/10.1016/S0947-3580(98)70113-X}
	{\bibfield  {journal} {\bibinfo  {journal} {Eur. J. Control}\ }\textbf
		{\bibinfo {volume} {4}},\ \bibinfo {pages} {176} (\bibinfo {year}
		{1998})}\BibitemShut {NoStop}%
	\bibitem [{\citenamefont {Pennestr{\`\i}}\ \emph {et~al.}(2016)\citenamefont
		{Pennestr{\`\i}}, \citenamefont {Rossi}, \citenamefont {Salvini},\ and\
		\citenamefont {Valentini}}]{pennestri2016review}%
	\BibitemOpen
	\bibfield  {author} {\bibinfo {author} {\bibfnamefont {E.}~\bibnamefont
			{Pennestr{\`\i}}}, \bibinfo {author} {\bibfnamefont {V.}~\bibnamefont
			{Rossi}}, \bibinfo {author} {\bibfnamefont {P.}~\bibnamefont {Salvini}},\
		and\ \bibinfo {author} {\bibfnamefont {P.~P.}\ \bibnamefont {Valentini}},\
	}\href {https://doi.org/10.1007/s11071-015-2485-3} {\bibfield  {journal}
		{\bibinfo  {journal} {Nonlinear Dyn.}\ }\textbf {\bibinfo {volume} {83}},\
		\bibinfo {pages} {1785} (\bibinfo {year} {2016})}\BibitemShut {NoStop}%
	\bibitem [{\citenamefont {de~Gennes}(2005)}]{deGennes:2005}%
	\BibitemOpen
	\bibfield  {author} {\bibinfo {author} {\bibfnamefont {P.-G.}\ \bibnamefont
			{de~Gennes}},\ }\href {https://doi.org/10.1007/s10955-005-4650-4} {\bibfield
		{journal} {\bibinfo  {journal} {J. Stat. Phys.}\ }\textbf {\bibinfo {volume}
			{119}},\ \bibinfo {pages} {953} (\bibinfo {year} {2005})}\BibitemShut
	{NoStop}%
	\bibitem [{\citenamefont {Hayakawa}(2005)}]{hayakawa2005langevin}%
	\BibitemOpen
	\bibfield  {author} {\bibinfo {author} {\bibfnamefont {H.}~\bibnamefont
			{Hayakawa}},\ }\href {https://doi.org/10.1016/j.physd.2004.12.011} {\bibfield
		{journal} {\bibinfo  {journal} {Physica D}\ }\textbf {\bibinfo {volume}
			{205}},\ \bibinfo {pages} {48} (\bibinfo {year} {2005})}\BibitemShut
	{NoStop}%
	\bibitem [{\citenamefont {Daniel}\ \emph {et~al.}(2005)\citenamefont {Daniel},
		\citenamefont {Chaudhury},\ and\ \citenamefont {de~Gennes}}]{Daniel2005}%
	\BibitemOpen
	\bibfield  {author} {\bibinfo {author} {\bibfnamefont {S.}~\bibnamefont
			{Daniel}}, \bibinfo {author} {\bibfnamefont {M.~K.}\ \bibnamefont
			{Chaudhury}},\ and\ \bibinfo {author} {\bibfnamefont {P.-G.}\ \bibnamefont
			{de~Gennes}},\ }\href {https://doi.org/10.1021/la046886s} {\bibfield
		{journal} {\bibinfo  {journal} {Langmuir}\ }\textbf {\bibinfo {volume}
			{21}},\ \bibinfo {pages} {4240} (\bibinfo {year} {2005})}\BibitemShut
	{NoStop}%
	\bibitem [{\citenamefont {Touchette}\ \emph {et~al.}(2010)\citenamefont
		{Touchette}, \citenamefont {der Straeten},\ and\ \citenamefont
		{Just}}]{Touchette/etal:2010}%
	\BibitemOpen
	\bibfield  {author} {\bibinfo {author} {\bibfnamefont {H.}~\bibnamefont
			{Touchette}}, \bibinfo {author} {\bibfnamefont {E.~V.}\ \bibnamefont {der
				Straeten}},\ and\ \bibinfo {author} {\bibfnamefont {W.}~\bibnamefont
			{Just}},\ }\href {https://doi.org/10.1088/1751-8113/43/44/445002} {\bibfield
		{journal} {\bibinfo  {journal} {J. Phys. A: Math. Theor.}\ }\textbf {\bibinfo
			{volume} {43}},\ \bibinfo {pages} {445002} (\bibinfo {year}
		{2010})}\BibitemShut {NoStop}%
	\bibitem [{\citenamefont {Chen}\ and\ \citenamefont
		{Just}(2014{\natexlab{a}})}]{chen2014first}%
	\BibitemOpen
	\bibfield  {author} {\bibinfo {author} {\bibfnamefont {Y.}~\bibnamefont
			{Chen}}\ and\ \bibinfo {author} {\bibfnamefont {W.}~\bibnamefont {Just}},\
	}\href {https://doi.org/10.1103/PhysRevE.89.022103} {\bibfield  {journal}
		{\bibinfo  {journal} {Phys. Rev. E}\ }\textbf {\bibinfo {volume} {89}},\
		\bibinfo {pages} {022103} (\bibinfo {year} {2014}{\natexlab{a}})}\BibitemShut
	{NoStop}%
	\bibitem [{\citenamefont {Lequy}\ and\ \citenamefont
		{Menzel}(2023)}]{lequy2023stochastic}%
	\BibitemOpen
	\bibfield  {author} {\bibinfo {author} {\bibfnamefont {T.}~\bibnamefont
			{Lequy}}\ and\ \bibinfo {author} {\bibfnamefont {A.~M.}\ \bibnamefont
			{Menzel}},\ }\href {https://doi.org/10.1088/1742-5468/aa8c1f} {\bibfield
		{journal} {\bibinfo  {journal} {Phys. Rev. E}\ }\textbf {\bibinfo {volume}
			{108}},\ \bibinfo {pages} {064606} (\bibinfo {year} {2023})}\BibitemShut
	{NoStop}%
	\bibitem [{\citenamefont {Plati}\ \emph {et~al.}(2023)\citenamefont {Plati},
		\citenamefont {Puglisi},\ and\ \citenamefont
		{Sarracino}}]{plati2023thermodynamic}%
	\BibitemOpen
	\bibfield  {author} {\bibinfo {author} {\bibfnamefont {A.}~\bibnamefont
			{Plati}}, \bibinfo {author} {\bibfnamefont {A.}~\bibnamefont {Puglisi}},\
		and\ \bibinfo {author} {\bibfnamefont {A.}~\bibnamefont {Sarracino}},\ }\href
	{https://doi.org/10.1088/1751-8121/ad358d} {\bibfield  {journal} {\bibinfo  {journal}
			{J. Phys. A: Math. Theor.}\ }\textbf {\bibinfo {volume}
			{57}},\ \bibinfo {pages} {155001}  (\bibinfo {year} {2024})}\BibitemShut {NoStop}%
	\bibitem [{\citenamefont {Gnoli}\ \emph {et~al.}(2013)\citenamefont {Gnoli},
		\citenamefont {Puglisi},\ and\ \citenamefont
		{Touchette}}]{gnoli2013granular}%
	\BibitemOpen
	\bibfield  {author} {\bibinfo {author} {\bibfnamefont {A.}~\bibnamefont
			{Gnoli}}, \bibinfo {author} {\bibfnamefont {A.}~\bibnamefont {Puglisi}},\
		and\ \bibinfo {author} {\bibfnamefont {H.}~\bibnamefont {Touchette}},\ }\href
	{https://doi.org/10.1209/0295-5075/102/14002} {\bibfield  {journal} {\bibinfo
			{journal} {Europhys. Lett.}\ }\textbf {\bibinfo {volume} {102}},\ \bibinfo
		{pages} {14002} (\bibinfo {year} {2013})}\BibitemShut {NoStop}%
	\bibitem [{\citenamefont {Lema{\^\i}tre}\ \emph {et~al.}(2021)\citenamefont
		{Lema{\^\i}tre}, \citenamefont {Mondal}, \citenamefont {Procaccia},\ and\
		\citenamefont {Roy}}]{lemaitre2021stress}%
	\BibitemOpen
	\bibfield  {author} {\bibinfo {author} {\bibfnamefont {A.}~\bibnamefont
			{Lema{\^\i}tre}}, \bibinfo {author} {\bibfnamefont {C.}~\bibnamefont
			{Mondal}}, \bibinfo {author} {\bibfnamefont {I.}~\bibnamefont {Procaccia}},\
		and\ \bibinfo {author} {\bibfnamefont {S.}~\bibnamefont {Roy}},\ }\href
	{https://doi.org/10.1103/PhysRevB.103.054110} {\bibfield  {journal} {\bibinfo
			{journal} {Phys. Rev. B}\ }\textbf {\bibinfo {volume} {103}},\ \bibinfo
		{pages} {054110} (\bibinfo {year} {2021})}\BibitemShut {NoStop}%
	\bibitem [{\citenamefont {Baule}\ and\ \citenamefont
		{Sollich}(2012)}]{baule2012singular}%
	\BibitemOpen
	\bibfield  {author} {\bibinfo {author} {\bibfnamefont {A.}~\bibnamefont
			{Baule}}\ and\ \bibinfo {author} {\bibfnamefont {P.}~\bibnamefont
			{Sollich}},\ }\href {https://doi.org/10.1209/0295-5075/97/20001} {\bibfield
		{journal} {\bibinfo  {journal} {Europhys. Lett.}\ }\textbf {\bibinfo {volume}
			{97}},\ \bibinfo {pages} {20001} (\bibinfo {year} {2012})}\BibitemShut
	{NoStop}%
	\bibitem [{\citenamefont {Chen}\ and\ \citenamefont
		{Just}(2014{\natexlab{b}})}]{chen2014large}%
	\BibitemOpen
	\bibfield  {author} {\bibinfo {author} {\bibfnamefont {Y.}~\bibnamefont
			{Chen}}\ and\ \bibinfo {author} {\bibfnamefont {W.}~\bibnamefont {Just}},\
	}\href {https://doi.org/10.1103/PhysRevE.90.042102} {\bibfield  {journal}
		{\bibinfo  {journal} {Phys. Rev. E}\ }\textbf {\bibinfo {volume} {90}},\
		\bibinfo {pages} {042102} (\bibinfo {year} {2014}{\natexlab{b}})}\BibitemShut
	{NoStop}%
	\bibitem [{\citenamefont {Manacorda}\ \emph {et~al.}(2014)\citenamefont
		{Manacorda}, \citenamefont {Puglisi},\ and\ \citenamefont
		{Sarracino}}]{manacorda2014coulomb}%
	\BibitemOpen
	\bibfield  {author} {\bibinfo {author} {\bibfnamefont {A.}~\bibnamefont
			{Manacorda}}, \bibinfo {author} {\bibfnamefont {A.}~\bibnamefont {Puglisi}},\
		and\ \bibinfo {author} {\bibfnamefont {A.}~\bibnamefont {Sarracino}},\ }\href
	{https://doi.org/10.1088/0253-6102/62/4/08} {\bibfield  {journal} {\bibinfo
			{journal} {Commun. Theor. Phys.}\ }\textbf {\bibinfo {volume} {62}},\
		\bibinfo {pages} {505} (\bibinfo {year} {2014})}\BibitemShut {NoStop}%
	\bibitem [{\citenamefont {Semeraro}\ \emph {et~al.}(2023)\citenamefont
		{Semeraro}, \citenamefont {Gonnella}, \citenamefont {Lippiello},\ and\
		\citenamefont {Sarracino}}]{semeraro2023diffusion}%
	\BibitemOpen
	\bibfield  {author} {\bibinfo {author} {\bibfnamefont {M.}~\bibnamefont
			{Semeraro}}, \bibinfo {author} {\bibfnamefont {G.}~\bibnamefont {Gonnella}},
		\bibinfo {author} {\bibfnamefont {E.}~\bibnamefont {Lippiello}},\ and\
		\bibinfo {author} {\bibfnamefont {A.}~\bibnamefont {Sarracino}},\ }\href
	{https://doi.org/10.3390/sym15010200} {\bibfield  {journal} {\bibinfo
			{journal} {Symmetry}\ }\textbf {\bibinfo {volume} {15}},\ \bibinfo {pages}
		{200} (\bibinfo {year} {2023})}\BibitemShut {NoStop}%
	\bibitem [{\citenamefont {Sarracino}\ \emph {et~al.}(2013)\citenamefont
		{Sarracino}, \citenamefont {Gnoli},\ and\ \citenamefont
		{Puglisi}}]{sarracino2013ratchet}%
	\BibitemOpen
	\bibfield  {author} {\bibinfo {author} {\bibfnamefont {A.}~\bibnamefont
			{Sarracino}}, \bibinfo {author} {\bibfnamefont {A.}~\bibnamefont {Gnoli}},\
		and\ \bibinfo {author} {\bibfnamefont {A.}~\bibnamefont {Puglisi}},\ }\href
	{https://doi.org/10.1103/PhysRevE.87.040101(R)} {\bibfield  {journal} {\bibinfo
			{journal} {Phys. Rev. E}\ }\textbf {\bibinfo {volume} {87}},\ \bibinfo
		{pages} {040101} (\bibinfo {year} {2013})}\BibitemShut {NoStop}%
	\bibitem [{\citenamefont {Sano}\ and\ \citenamefont
		{Hayakawa}(2014)}]{sano2014roles}%
	\BibitemOpen
	\bibfield  {author} {\bibinfo {author} {\bibfnamefont {T.~G.}\ \bibnamefont
			{Sano}}\ and\ \bibinfo {author} {\bibfnamefont {H.}~\bibnamefont
			{Hayakawa}},\ }\href {https://doi.org/10.1103/PhysRevE.89.032104} {\bibfield
		{journal} {\bibinfo  {journal} {Phys. Rev. E}\ }\textbf {\bibinfo {volume}
			{89}},\ \bibinfo {pages} {032104} (\bibinfo {year} {2014})}\BibitemShut
	{NoStop}%
	\bibitem [{\citenamefont {Marchetti}\ \emph {et~al.}(2013)\citenamefont
		{Marchetti}, \citenamefont {Joanny}, \citenamefont {Ramaswamy}, \citenamefont
		{Liverpool}, \citenamefont {Prost}, \citenamefont {Rao},\ and\ \citenamefont
		{Simha}}]{marchetti2013hydrodynamics}%
	\BibitemOpen
	\bibfield  {author} {\bibinfo {author} {\bibfnamefont {M.~C.}~\bibnamefont
			{Marchetti}}, \bibinfo {author} {\bibfnamefont {J.~F.}~\bibnamefont {Joanny}},
		\bibinfo {author} {\bibfnamefont {S.}~\bibnamefont {Ramaswamy}}, \bibinfo
		{author} {\bibfnamefont {T.~B.}~\bibnamefont {Liverpool}}, \bibinfo {author}
		{\bibfnamefont {J.}~\bibnamefont {Prost}}, \bibinfo {author} {\bibfnamefont
			{M.}~\bibnamefont {Rao}},\ and\ \bibinfo {author} {\bibfnamefont {R.~A.}\
			\bibnamefont {Simha}},\ }\href {https://doi.org/10.1103/RevModPhys.85.1143}
	{\bibfield  {journal} {\bibinfo  {journal} {Rev. Mod. Phys.}\ }\textbf
		{\bibinfo {volume} {85}},\ \bibinfo {pages} {1143} (\bibinfo {year}
		{2013})}\BibitemShut {NoStop}%
	\bibitem [{\citenamefont {Elgeti}\ \emph
		{et~al.}(2015{\natexlab{a}})\citenamefont {Elgeti}, \citenamefont {Winkler},\
		and\ \citenamefont {Gompper}}]{Elgeti2015}%
	\BibitemOpen
	\bibfield  {author} {\bibinfo {author} {\bibfnamefont {J.}~\bibnamefont
			{Elgeti}}, \bibinfo {author} {\bibfnamefont {R.~G.}\ \bibnamefont
			{Winkler}},\ and\ \bibinfo {author} {\bibfnamefont {G.}~\bibnamefont
			{Gompper}},\ }\href {https://doi.org/10.1088/0034-4885/78/5/056601}
	{\bibfield  {journal} {\bibinfo  {journal} {Rep. Prog. Phys.}\ }\textbf
		{\bibinfo {volume} {78}},\ \bibinfo {pages} {56601} (\bibinfo {year}
		{2015}{\natexlab{a}})}\BibitemShut {NoStop}%
	\bibitem [{\citenamefont {Bechinger}\ \emph {et~al.}(2016)\citenamefont
		{Bechinger}, \citenamefont {Di~Leonardo}, \citenamefont {L{\"o}wen},
		\citenamefont {Reichhardt}, \citenamefont {Volpe},\ and\ \citenamefont
		{Volpe}}]{bechinger2016active}%
	\BibitemOpen
	\bibfield  {author} {\bibinfo {author} {\bibfnamefont {C.}~\bibnamefont
			{Bechinger}}, \bibinfo {author} {\bibfnamefont {R.}~\bibnamefont
			{Di~Leonardo}}, \bibinfo {author} {\bibfnamefont {H.}~\bibnamefont
			{L{\"o}wen}}, \bibinfo {author} {\bibfnamefont {C.}~\bibnamefont
			{Reichhardt}}, \bibinfo {author} {\bibfnamefont {G.}~\bibnamefont {Volpe}},\
		and\ \bibinfo {author} {\bibfnamefont {G.}~\bibnamefont {Volpe}},\ }\href
	{https://doi.org/10.1103/RevModPhys.88.045006} {\bibfield  {journal}
		{\bibinfo  {journal} {Rev. Mod. Phys.}\ }\textbf {\bibinfo {volume} {88}},\
		\bibinfo {pages} {045006} (\bibinfo {year} {2016})}\BibitemShut {NoStop}%
	\bibitem [{\citenamefont {O’Byrne}\ \emph {et~al.}(2022)\citenamefont
		{O’Byrne}, \citenamefont {Kafri}, \citenamefont {Tailleur},\ and\
		\citenamefont {van Wijland}}]{o2022time}%
	\BibitemOpen
	\bibfield  {author} {\bibinfo {author} {\bibfnamefont {J.}~\bibnamefont
			{O’Byrne}}, \bibinfo {author} {\bibfnamefont {Y.}~\bibnamefont {Kafri}},
		\bibinfo {author} {\bibfnamefont {J.}~\bibnamefont {Tailleur}},\ and\
		\bibinfo {author} {\bibfnamefont {F.}~\bibnamefont {van Wijland}},\ }\href
	{https://doi.org/10.1038/s42254-021-00406-2} {\bibfield  {journal} {\bibinfo
			{journal} {Nat. Rev. Phys.}\ }\textbf {\bibinfo {volume} {4}},\ \bibinfo
		{pages} {167} (\bibinfo {year} {2022})}\BibitemShut {NoStop}%
	\bibitem [{\citenamefont {Fodor}\ \emph {et~al.}(2021)\citenamefont {Fodor},
		\citenamefont {Jack},\ and\ \citenamefont
		{Cates}}]{fodor2021irreversibility}%
	\BibitemOpen
	\bibfield  {author} {\bibinfo {author} {\bibfnamefont {{\'E}.}~\bibnamefont
			{Fodor}}, \bibinfo {author} {\bibfnamefont {R.~L.}\ \bibnamefont {Jack}},\
		and\ \bibinfo {author} {\bibfnamefont {M.~E.}\ \bibnamefont {Cates}},\ }\href
	{https://doi.org/10.1146/annurev-conmatphys-031720-032419} {\bibfield
		{journal} {\bibinfo  {journal} {Annu. Rev. Condens. Matter Phys.}\ }\textbf
		{\bibinfo {volume} {13}},\ \bibinfo {pages} {215} (\bibinfo {year}
		{2021})}\BibitemShut {NoStop}%
	\bibitem [{\citenamefont {Z\"ottl}\ and\ \citenamefont
		{Stark}(2016)}]{Zoettl/Stark:2016}%
	\BibitemOpen
	\bibfield  {author} {\bibinfo {author} {\bibfnamefont {A.}~\bibnamefont
			{Z\"ottl}}\ and\ \bibinfo {author} {\bibfnamefont {H.}~\bibnamefont
			{Stark}},\ }\href {https://doi.org/10.1088/0953-8984/28/25/253001} {\bibfield
		{journal} {\bibinfo  {journal} {J. Phys.: Condens. Matter}\ }\textbf
		{\bibinfo {volume} {28}},\ \bibinfo {pages} {253001} (\bibinfo {year}
		{2016})}\BibitemShut {NoStop}%
	\bibitem [{\citenamefont {L{\"o}wen}(2020)}]{lowen2020inertial}%
	\BibitemOpen
	\bibfield  {author} {\bibinfo {author} {\bibfnamefont {H.}~\bibnamefont
			{L{\"o}wen}},\ }\href {https://doi.org/10.1063/1.5134455} {\bibfield
		{journal} {\bibinfo  {journal} {J. Chem. Phys.}\ }\textbf {\bibinfo {volume}
			{152}},\ \bibinfo {pages} {040901} (\bibinfo {year} {2020})}\BibitemShut
	{NoStop}%
	\bibitem [{\citenamefont {Romanczuk}\ and\ \citenamefont
		{Schimansky-Geier}(2011)}]{Romanczuk/Schimansky:2011}%
	\BibitemOpen
	\bibfield  {author} {\bibinfo {author} {\bibfnamefont {P.}~\bibnamefont
			{Romanczuk}}\ and\ \bibinfo {author} {\bibfnamefont {L.}~\bibnamefont
			{Schimansky-Geier}},\ }\href {https://doi.org/10.1103/PhysRevLett.106.230601}
	{\bibfield  {journal} {\bibinfo  {journal} {Phys. Rev. Lett.}\ }\textbf
		{\bibinfo {volume} {106}},\ \bibinfo {pages} {230601} (\bibinfo {year}
		{2011})}\BibitemShut {NoStop}%
	\bibitem [{\citenamefont {Romanczuk}\ \emph {et~al.}(2012)\citenamefont
		{Romanczuk}, \citenamefont {B{\"a}r}, \citenamefont {Ebeling}, \citenamefont
		{Lindner},\ and\ \citenamefont {Schimansky-Geier}}]{Romanczuk2012}%
	\BibitemOpen
	\bibfield  {author} {\bibinfo {author} {\bibfnamefont {P.}~\bibnamefont
			{Romanczuk}}, \bibinfo {author} {\bibfnamefont {M.}~\bibnamefont {B{\"a}r}},
		\bibinfo {author} {\bibfnamefont {W.}~\bibnamefont {Ebeling}}, \bibinfo
		{author} {\bibfnamefont {B.}~\bibnamefont {Lindner}},\ and\ \bibinfo {author}
		{\bibfnamefont {L.}~\bibnamefont {Schimansky-Geier}},\ }\href
	{https://doi.org/10.1140/epjst/e2012-01529-y} {\bibfield  {journal} {\bibinfo
			{journal} {Eur. Phys. J. Spec. Top.}\ }\textbf {\bibinfo {volume} {202}},\
		\bibinfo {pages} {1} (\bibinfo {year} {2012})}\BibitemShut {NoStop}%
	\bibitem [{\citenamefont {Elgeti}\ \emph
		{et~al.}(2015{\natexlab{b}})\citenamefont {Elgeti}, \citenamefont {Winkler},\
		and\ \citenamefont {Gompper}}]{Elgeti_2015}%
	\BibitemOpen
	\bibfield  {author} {\bibinfo {author} {\bibfnamefont {J.}~\bibnamefont
			{Elgeti}}, \bibinfo {author} {\bibfnamefont {R.~G.}\ \bibnamefont
			{Winkler}},\ and\ \bibinfo {author} {\bibfnamefont {G.}~\bibnamefont
			{Gompper}},\ }\href {https://doi.org/10.1088/0034-4885/78/5/056601}
	{\bibfield  {journal} {\bibinfo  {journal} {Rep. Prog. Phys.}\ }\textbf
		{\bibinfo {volume} {78}},\ \bibinfo {pages} {056601} (\bibinfo {year}
		{2015}{\natexlab{b}})}\BibitemShut {NoStop}%
	\bibitem [{\citenamefont {Aranson}\ \emph {et~al.}(2007)\citenamefont
		{Aranson}, \citenamefont {Volfson},\ and\ \citenamefont
		{Tsimring}}]{aranson2007swirling}%
	\BibitemOpen
	\bibfield  {author} {\bibinfo {author} {\bibfnamefont {I.~S.}\ \bibnamefont
			{Aranson}}, \bibinfo {author} {\bibfnamefont {D.}~\bibnamefont {Volfson}},\
		and\ \bibinfo {author} {\bibfnamefont {L.~S.}\ \bibnamefont {Tsimring}},\
	}\href {https://doi.org/10.1103/PhysRevE.75.051301} {\bibfield  {journal}
		{\bibinfo  {journal} {Phys. Rev. E}\ }\textbf {\bibinfo {volume} {75}},\
		\bibinfo {pages} {051301} (\bibinfo {year} {2007})}\BibitemShut {NoStop}%
	\bibitem [{\citenamefont {Kudrolli}\ \emph {et~al.}(2008)\citenamefont
		{Kudrolli}, \citenamefont {Lumay}, \citenamefont {Volfson},\ and\
		\citenamefont {Tsimring}}]{kudrolli2008}%
	\BibitemOpen
	\bibfield  {author} {\bibinfo {author} {\bibfnamefont {A.}~\bibnamefont
			{Kudrolli}}, \bibinfo {author} {\bibfnamefont {G.}~\bibnamefont {Lumay}},
		\bibinfo {author} {\bibfnamefont {D.}~\bibnamefont {Volfson}},\ and\ \bibinfo
		{author} {\bibfnamefont {L.~S.}\ \bibnamefont {Tsimring}},\ }\href
	{https://doi.org/10.1103/PhysRevLett.100.058001} {\bibfield  {journal}
		{\bibinfo  {journal} {Phys. Rev. Lett.}\ }\textbf {\bibinfo {volume} {100}},\
		\bibinfo {pages} {058001} (\bibinfo {year} {2008})}\BibitemShut {NoStop}%
	\bibitem [{\citenamefont {Kumar}\ \emph {et~al.}(2014)\citenamefont {Kumar},
		\citenamefont {Soni}, \citenamefont {Ramaswamy},\ and\ \citenamefont
		{Sood}}]{kumar2014flocking}%
	\BibitemOpen
	\bibfield  {author} {\bibinfo {author} {\bibfnamefont {N.}~\bibnamefont
			{Kumar}}, \bibinfo {author} {\bibfnamefont {H.}~\bibnamefont {Soni}},
		\bibinfo {author} {\bibfnamefont {S.}~\bibnamefont {Ramaswamy}},\ and\
		\bibinfo {author} {\bibfnamefont {A.}~\bibnamefont {Sood}},\ }\href
	{https://doi.org/10.1038/ncomms5688} {\bibfield  {journal} {\bibinfo
			{journal} {Nat. Commun.}\ }\textbf {\bibinfo {volume} {5}},\ \bibinfo {pages}
		{4688} (\bibinfo {year} {2014})}\BibitemShut {NoStop}%
	\bibitem [{\citenamefont {Koumakis}\ \emph {et~al.}(2016)\citenamefont
		{Koumakis}, \citenamefont {Gnoli}, \citenamefont {Maggi}, \citenamefont
		{Puglisi},\ and\ \citenamefont {Di~Leonardo}}]{Koumakis2016}%
	\BibitemOpen
	\bibfield  {author} {\bibinfo {author} {\bibfnamefont {N.}~\bibnamefont
			{Koumakis}}, \bibinfo {author} {\bibfnamefont {A.}~\bibnamefont {Gnoli}},
		\bibinfo {author} {\bibfnamefont {C.}~\bibnamefont {Maggi}}, \bibinfo
		{author} {\bibfnamefont {A.}~\bibnamefont {Puglisi}},\ and\ \bibinfo {author}
		{\bibfnamefont {R.}~\bibnamefont {Di~Leonardo}},\ }\href
	{https://doi.org/10.1088/1367-2630/18/11/113046} {\bibfield  {journal}
		{\bibinfo  {journal} {New J. Phys.}\ }\textbf {\bibinfo {volume} {18}},\
		\bibinfo {pages} {113046} (\bibinfo {year} {2016})}\BibitemShut {NoStop}%
	\bibitem [{\citenamefont {Agrawal}\ and\ \citenamefont
		{Glotzer}(2020)}]{Agrawal2020}%
	\BibitemOpen
	\bibfield  {author} {\bibinfo {author} {\bibfnamefont {M.}~\bibnamefont
			{Agrawal}}\ and\ \bibinfo {author} {\bibfnamefont {S.~C.}\ \bibnamefont
			{Glotzer}},\ }\href {https://doi.org/10.1073/pnas.1922635117} {\bibfield
		{journal} {\bibinfo  {journal} {Proc. Natl. Acad. Sci. U.S.A.}\ }\textbf
		{\bibinfo {volume} {117}},\ \bibinfo {pages} {8700} (\bibinfo {year}
		{2020})}\BibitemShut {NoStop}%
	\bibitem [{\citenamefont {Walsh}\ \emph {et~al.}(2017)\citenamefont {Walsh},
		\citenamefont {Wagner}, \citenamefont {Schlossberg}, \citenamefont {Olson},
		\citenamefont {Baskaran},\ and\ \citenamefont {Menon}}]{walsh2017noise}%
	\BibitemOpen
	\bibfield  {author} {\bibinfo {author} {\bibfnamefont {L.}~\bibnamefont
			{Walsh}}, \bibinfo {author} {\bibfnamefont {C.~G.}\ \bibnamefont {Wagner}},
		\bibinfo {author} {\bibfnamefont {S.}~\bibnamefont {Schlossberg}}, \bibinfo
		{author} {\bibfnamefont {C.}~\bibnamefont {Olson}}, \bibinfo {author}
		{\bibfnamefont {A.}~\bibnamefont {Baskaran}},\ and\ \bibinfo {author}
		{\bibfnamefont {N.}~\bibnamefont {Menon}},\ }\href
	{https://doi.org/10.1039/C7SM01206C} {\bibfield  {journal} {\bibinfo
			{journal} {Soft~Matter}\ }\textbf {\bibinfo {volume} {13}},\ \bibinfo {pages}
		{8964} (\bibinfo {year} {2017})}\BibitemShut {NoStop}%
	\bibitem [{\citenamefont {Baconnier}\ \emph {et~al.}(2022)\citenamefont
		{Baconnier}, \citenamefont {Shohat}, \citenamefont {L{\'o}pez}, \citenamefont
		{Coulais}, \citenamefont {D{\'e}mery}, \citenamefont {D{\"u}ring},\ and\
		\citenamefont {Dauchot}}]{baconnier2022selective}%
	\BibitemOpen
	\bibfield  {author} {\bibinfo {author} {\bibfnamefont {P.}~\bibnamefont
			{Baconnier}}, \bibinfo {author} {\bibfnamefont {D.}~\bibnamefont {Shohat}},
		\bibinfo {author} {\bibfnamefont {C.~H.}\ \bibnamefont {L{\'o}pez}}, \bibinfo
		{author} {\bibfnamefont {C.}~\bibnamefont {Coulais}}, \bibinfo {author}
		{\bibfnamefont {V.}~\bibnamefont {D{\'e}mery}}, \bibinfo {author}
		{\bibfnamefont {G.}~\bibnamefont {D{\"u}ring}},\ and\ \bibinfo {author}
		{\bibfnamefont {O.}~\bibnamefont {Dauchot}},\ }\href
	{https://doi.org/10.1038/s41567-022-01704-x} {\bibfield  {journal} {\bibinfo
			{journal} {Nat. Phys.}\ }\textbf {\bibinfo {volume} {18}},\ \bibinfo {pages}
		{1234} (\bibinfo {year} {2022})}\BibitemShut {NoStop}%
	\bibitem [{\citenamefont {Caprini}\ \emph
		{et~al.}(2024{\natexlab{a}})\citenamefont {Caprini}, \citenamefont {Breoni},
		\citenamefont {Ldov}, \citenamefont {Scholz},\ and\ \citenamefont
		{L{\"o}wen}}]{caprini2024dynamical}%
	\BibitemOpen
	\bibfield  {author} {\bibinfo {author} {\bibfnamefont {L.}~\bibnamefont
			{Caprini}}, \bibinfo {author} {\bibfnamefont {D.}~\bibnamefont {Breoni}},
		\bibinfo {author} {\bibfnamefont {A.}~\bibnamefont {Ldov}}, \bibinfo {author}
		{\bibfnamefont {C.}~\bibnamefont {Scholz}},\ and\ \bibinfo {author}
		{\bibfnamefont {H.}~\bibnamefont {L{\"o}wen}},\ }\href
	{https://doi.org/10.1038/s42005-024-01835-y} {\bibfield  {journal} {\bibinfo  {journal}
			{Commun. Phys.}} 
		\textbf {\bibinfo {volume} {7}},\ \bibinfo {pages}
		{343} (\bibinfo {year} {2024}{\natexlab{a}})}\BibitemShut
	{NoStop}%
	\bibitem [{\citenamefont {Scholz}\ \emph {et~al.}(2018)\citenamefont {Scholz},
		\citenamefont {Jahanshahi}, \citenamefont {Ldov},\ and\ \citenamefont
		{L{\"o}wen}}]{scholz2018inertial}%
	\BibitemOpen
	\bibfield  {author} {\bibinfo {author} {\bibfnamefont {C.}~\bibnamefont
			{Scholz}}, \bibinfo {author} {\bibfnamefont {S.}~\bibnamefont {Jahanshahi}},
		\bibinfo {author} {\bibfnamefont {A.}~\bibnamefont {Ldov}},\ and\ \bibinfo
		{author} {\bibfnamefont {H.}~\bibnamefont {L{\"o}wen}},\ }\href
	{https://doi.org/10.1038/s41467-018-07596-x} {\bibfield  {journal} {\bibinfo
			{journal} {Nat. Commun.}\ }\textbf {\bibinfo {volume} {9}},\ \bibinfo {pages}
		{5156} (\bibinfo {year} {2018})}\BibitemShut {NoStop}%
	\bibitem [{Note1()}]{Note1}%
	\BibitemOpen
	\bibinfo {note} {See Supplemental Material, which includes Refs.~\cite {scholz2018inertial, Caroli1981,
			Maslov1981, Antonov2023, Landau}, for details on the analytical
		calculations.}\BibitemShut {NoStop}%
	\bibitem [{\citenamefont {Caprini}\ \emph
		{et~al.}(2024{\natexlab{b}})\citenamefont {Caprini}, \citenamefont {Ldov},
		\citenamefont {Gupta}, \citenamefont {Ellenberg}, \citenamefont {Wittmann},
		\citenamefont {L{\"o}wen},\ and\ \citenamefont
		{Scholz}}]{caprini2024emergent}%
	\BibitemOpen
	\bibfield  {author} {\bibinfo {author} {\bibfnamefont {L.}~\bibnamefont
			{Caprini}}, \bibinfo {author} {\bibfnamefont {A.}~\bibnamefont {Ldov}},
		\bibinfo {author} {\bibfnamefont {R.~K.}\ \bibnamefont {Gupta}}, \bibinfo
		{author} {\bibfnamefont {H.}~\bibnamefont {Ellenberg}}, \bibinfo {author}
		{\bibfnamefont {R.}~\bibnamefont {Wittmann}}, \bibinfo {author}
		{\bibfnamefont {H.}~\bibnamefont {L{\"o}wen}},\ and\ \bibinfo {author}
		{\bibfnamefont {C.}~\bibnamefont {Scholz}},\ }\href
	{https://doi.org/10.1038/s42005-024-01540-w} {\bibfield  {journal} {\bibinfo
			{journal} {Commun. Phys.}\ }\textbf {\bibinfo {volume} {7}},\ \bibinfo
		{pages} {52} (\bibinfo {year} {2024}{\natexlab{b}})}\BibitemShut {NoStop}%
	\bibitem [{\citenamefont {Marton}\ and\ \citenamefont
		{Lantos}(2007)}]{Lrinc/Bla:2007}%
	\BibitemOpen
	\bibfield  {author} {\bibinfo {author} {\bibfnamefont {L.}~\bibnamefont
			{Marton}}\ and\ \bibinfo {author} {\bibfnamefont {B.}~\bibnamefont
			{Lantos}},\ }\href {https://doi.org/10.1109/TIE.2006.888804} {\bibfield
		{journal} {\bibinfo  {journal} {IEEE Trans. Ind. Electron.}\ }\textbf
		{\bibinfo {volume} {54}},\ \bibinfo {pages} {511} (\bibinfo {year}
		{2007})}\BibitemShut {NoStop}%
	\bibitem [{\citenamefont {Stribeck}(1902)}]{Stribeck:1902}%
	\BibitemOpen
	\bibfield  {author} {\bibinfo {author} {\bibfnamefont {R.}~\bibnamefont
			{Stribeck}},\ }\href@noop {} {\bibfield  {journal} {\bibinfo  {journal}
			{Z. Ver. Dtsch. Ing.}\ }\textbf {\bibinfo {volume}
			{46}},\ \bibinfo {pages} {1341} (\bibinfo {year} {1902})}\BibitemShut
	{NoStop}%
	\bibitem [{\citenamefont {Geffert}\ and\ \citenamefont
		{Just}(2017)}]{Geffert/Just:2017}%
	\BibitemOpen
	\bibfield  {author} {\bibinfo {author} {\bibfnamefont {P.~M.}\ \bibnamefont
			{Geffert}}\ and\ \bibinfo {author} {\bibfnamefont {W.}~\bibnamefont {Just}},\
	}\href {https://doi.org/10.1103/PhysRevE.95.062111} {\bibfield  {journal}
		{\bibinfo  {journal} {Phys. Rev. E}\ }\textbf {\bibinfo {volume} {95}},\
		\bibinfo {pages} {062111} (\bibinfo {year} {2017})}\BibitemShut {NoStop}%
	\bibitem [{\citenamefont {Szamel}(2014)}]{szamel2014self}%
	\BibitemOpen
	\bibfield  {author} {\bibinfo {author} {\bibfnamefont {G.}~\bibnamefont
			{Szamel}},\ }\href {https://doi.org/10.1103/PhysRevE.90.012111} {\bibfield
		{journal} {\bibinfo  {journal} {Phys. Rev. E}\ }\textbf {\bibinfo {volume}
			{90}},\ \bibinfo {pages} {012111} (\bibinfo {year} {2014})}\BibitemShut
	{NoStop}%
	\bibitem [{\citenamefont {Maggi}\ \emph {et~al.}(2014)\citenamefont {Maggi},
		\citenamefont {Paoluzzi}, \citenamefont {Pellicciotta}, \citenamefont
		{Lepore}, \citenamefont {Angelani},\ and\ \citenamefont
		{Di~Leonardo}}]{maggi2014generalized}%
	\BibitemOpen
	\bibfield  {author} {\bibinfo {author} {\bibfnamefont {C.}~\bibnamefont
			{Maggi}}, \bibinfo {author} {\bibfnamefont {M.}~\bibnamefont {Paoluzzi}},
		\bibinfo {author} {\bibfnamefont {N.}~\bibnamefont {Pellicciotta}}, \bibinfo
		{author} {\bibfnamefont {A.}~\bibnamefont {Lepore}}, \bibinfo {author}
		{\bibfnamefont {L.}~\bibnamefont {Angelani}},\ and\ \bibinfo {author}
		{\bibfnamefont {R.}~\bibnamefont {Di~Leonardo}},\ }\href
	{https://doi.org/10.1103/PhysRevLett.113.238303} {\bibfield  {journal}
		{\bibinfo  {journal} {Phys. Rev. Lett.}\ }\textbf {\bibinfo {volume} {113}},\
		\bibinfo {pages} {238303} (\bibinfo {year} {2014})}\BibitemShut {NoStop}%
	\bibitem [{\citenamefont {Wittmann}\ \emph {et~al.}(2017)\citenamefont
		{Wittmann}, \citenamefont {Maggi}, \citenamefont {Sharma}, \citenamefont
		{Scacchi}, \citenamefont {Brader},\ and\ \citenamefont
		{Marconi}}]{wittmann2017effective}%
	\BibitemOpen
	\bibfield  {author} {\bibinfo {author} {\bibfnamefont {R.}~\bibnamefont
			{Wittmann}}, \bibinfo {author} {\bibfnamefont {C.}~\bibnamefont {Maggi}},
		\bibinfo {author} {\bibfnamefont {A.}~\bibnamefont {Sharma}}, \bibinfo
		{author} {\bibfnamefont {A.}~\bibnamefont {Scacchi}}, \bibinfo {author}
		{\bibfnamefont {J.~M.}\ \bibnamefont {Brader}},\ and\ \bibinfo {author}
		{\bibfnamefont {U.~M.~B.}\ \bibnamefont {Marconi}},\ }\href
	{https://doi.org/10.1088/1742-5468/aa8c1f} {\bibfield  {journal} {\bibinfo
			{journal} {J. Stat. Mech.: Theory Exp.}\ }\textbf {\bibinfo {volume}
			{2017}}\bibinfo  {number} { (11)},\ \bibinfo {pages} {113207}}\BibitemShut
	{NoStop}%
	\bibitem [{\citenamefont {Caprini}\ and\ \citenamefont
		{Marconi}(2018)}]{caprini2018active}%
	\BibitemOpen
	\bibfield  {number} {  }\bibfield  {author} {\bibinfo {author} {\bibfnamefont
			{L.}~\bibnamefont {Caprini}}\ and\ \bibinfo {author} {\bibfnamefont
			{U.~M.~B.}\ \bibnamefont {Marconi}},\ }\href
	{https://doi.org/10.1039/C8SM01840E} {\bibfield  {journal} {\bibinfo
			{journal} {Soft~Matter}\ }\textbf {\bibinfo {volume} {14}},\ \bibinfo {pages}
		{9044} (\bibinfo {year} {2018})}\BibitemShut {NoStop}%
	\bibitem [{\citenamefont {Fily}(2019)}]{fily2019self}%
	\BibitemOpen
	\bibfield  {author} {\bibinfo {author} {\bibfnamefont {Y.}~\bibnamefont
			{Fily}},\ }\href {https://doi.org/10.1063/1.5085759} {\bibfield  {journal}
		{\bibinfo  {journal} {J. Chem. Phys.}\ }\textbf {\bibinfo {volume} {150}},\
		\bibinfo {pages} {174906} (\bibinfo {year} {2019})}\BibitemShut {NoStop}%
	\bibitem [{\citenamefont {Woillez}\ \emph {et~al.}(2020)\citenamefont
		{Woillez}, \citenamefont {Kafri},\ and\ \citenamefont
		{Gov}}]{woillez2020active}%
	\BibitemOpen
	\bibfield  {author} {\bibinfo {author} {\bibfnamefont {E.}~\bibnamefont
			{Woillez}}, \bibinfo {author} {\bibfnamefont {Y.}~\bibnamefont {Kafri}},\
		and\ \bibinfo {author} {\bibfnamefont {N.~S.}\ \bibnamefont {Gov}},\ }\href
	{https://doi.org/10.1103/PhysRevLett.124.118002} {\bibfield  {journal}
		{\bibinfo  {journal} {Phys. Rev. Lett.}\ }\textbf {\bibinfo {volume} {124}},\
		\bibinfo {pages} {118002} (\bibinfo {year} {2020})}\BibitemShut {NoStop}%
	\bibitem [{\citenamefont {Martin}\ \emph {et~al.}(2021)\citenamefont {Martin},
		\citenamefont {O'Byrne}, \citenamefont {Cates}, \citenamefont {Fodor},
		\citenamefont {Nardini}, \citenamefont {Tailleur},\ and\ \citenamefont
		{Van~Wijland}}]{martin2021statistical}%
	\BibitemOpen
	\bibfield  {author} {\bibinfo {author} {\bibfnamefont {D.}~\bibnamefont
			{Martin}}, \bibinfo {author} {\bibfnamefont {J.}~\bibnamefont {O'Byrne}},
		\bibinfo {author} {\bibfnamefont {M.~E.}\ \bibnamefont {Cates}}, \bibinfo
		{author} {\bibfnamefont {{\'E}.}~\bibnamefont {Fodor}}, \bibinfo {author}
		{\bibfnamefont {C.}~\bibnamefont {Nardini}}, \bibinfo {author} {\bibfnamefont
			{J.}~\bibnamefont {Tailleur}},\ and\ \bibinfo {author} {\bibfnamefont
			{F.}~\bibnamefont {Van~Wijland}},\ }\href
	{https://doi.org/10.1103/PhysRevE.103.032607} {\bibfield  {journal} {\bibinfo
			{journal} {Phys. Rev. E}\ }\textbf {\bibinfo {volume} {103}},\ \bibinfo
		{pages} {032607} (\bibinfo {year} {2021})}\BibitemShut {NoStop}%
	\bibitem [{\citenamefont {Keta}\ \emph {et~al.}(2022)\citenamefont {Keta},
		\citenamefont {Jack},\ and\ \citenamefont
		{Berthier}}]{PhysRevLett.129.048002}%
	\BibitemOpen
	\bibfield  {author} {\bibinfo {author} {\bibfnamefont {Y.-E.}\ \bibnamefont
			{Keta}}, \bibinfo {author} {\bibfnamefont {R.~L.}\ \bibnamefont {Jack}},\
		and\ \bibinfo {author} {\bibfnamefont {L.}~\bibnamefont {Berthier}},\ }\href
	{https://doi.org/10.1103/PhysRevLett.129.048002} {\bibfield  {journal}
		{\bibinfo  {journal} {Phys. Rev. Lett.}\ }\textbf {\bibinfo {volume} {129}},\
		\bibinfo {pages} {048002} (\bibinfo {year} {2022})}\BibitemShut {NoStop}%
	\bibitem [{\citenamefont {ten Hagen}\ \emph {et~al.}(2011)\citenamefont {ten
			Hagen}, \citenamefont {van Teeffelen},\ and\ \citenamefont
		{L{\"o}wen}}]{ten2011brownian}%
	\BibitemOpen
	\bibfield  {author} {\bibinfo {author} {\bibfnamefont {B.}~\bibnamefont {ten
				Hagen}}, \bibinfo {author} {\bibfnamefont {S.}~\bibnamefont {van
				Teeffelen}},\ and\ \bibinfo {author} {\bibfnamefont {H.}~\bibnamefont
			{L{\"o}wen}},\ }\href {https://doi.org/10.1088/0953-8984/23/19/194119}
	{\bibfield  {journal} {\bibinfo  {journal} {J. Phys.: Condens. Matter}\
		}\textbf {\bibinfo {volume} {23}},\ \bibinfo {pages} {194119} (\bibinfo
		{year} {2011})}\BibitemShut {NoStop}%
	\bibitem [{\citenamefont {Caprini}\ and\ \citenamefont {Marini
			Bettolo~Marconi}(2021)}]{caprini2021inertial}%
	\BibitemOpen
	\bibfield  {author} {\bibinfo {author} {\bibfnamefont {L.}~\bibnamefont
			{Caprini}}\ and\ \bibinfo {author} {\bibfnamefont {U.}~\bibnamefont {Marini
				Bettolo~Marconi}},\ }\href {https://doi.org/10.1063/5.0030940} {\bibfield
		{journal} {\bibinfo  {journal} {J. Chem. Phys.}\ }\textbf {\bibinfo {volume}
			{154}},\ \bibinfo {pages} {024902} (\bibinfo {year} {2021})}\BibitemShut
	{NoStop}%
	\bibitem [{\citenamefont {Caroli}\ \emph {et~al.}(1981)\citenamefont {Caroli},
		\citenamefont {Caroli},\ and\ \citenamefont {Roulet}}]{Caroli1981}%
	\BibitemOpen
	\bibfield  {author} {\bibinfo {author} {\bibfnamefont {B.}~\bibnamefont
			{Caroli}}, \bibinfo {author} {\bibfnamefont {C.}~\bibnamefont {Caroli}},\
		and\ \bibinfo {author} {\bibfnamefont {B.}~\bibnamefont {Roulet}},\ }\href
	{https://doi.org/10.1007/BF01106788} {\bibfield  {journal} {\bibinfo
			{journal} {J. Stat. Phys.}\ }\textbf {\bibinfo {volume} {26}},\ \bibinfo
		{pages} {83} (\bibinfo {year} {1981})}\BibitemShut {NoStop}%
	\bibitem [{\citenamefont {Dauchot}\ and\ \citenamefont
		{D{\'e}mery}(2019)}]{dauchot2019dynamics}%
	\BibitemOpen
	\bibfield  {author} {\bibinfo {author} {\bibfnamefont {O.}~\bibnamefont
			{Dauchot}}\ and\ \bibinfo {author} {\bibfnamefont {V.}~\bibnamefont
			{D{\'e}mery}},\ }\href {https://doi.org/10.1103/PhysRevLett.122.068002}
	{\bibfield  {journal} {\bibinfo  {journal} {Phys. Rev. Lett.}\ }\textbf
		{\bibinfo {volume} {122}},\ \bibinfo {pages} {068002} (\bibinfo {year}
		{2019})}\BibitemShut {NoStop}%
	\bibitem [{\citenamefont {Tapia-Ignacio}\ \emph {et~al.}(2021)\citenamefont
		{Tapia-Ignacio}, \citenamefont {Gutierrez-Martinez},\ and\ \citenamefont
		{Sandoval}}]{tapia2021trapped}%
	\BibitemOpen
	\bibfield  {author} {\bibinfo {author} {\bibfnamefont {C.}~\bibnamefont
			{Tapia-Ignacio}}, \bibinfo {author} {\bibfnamefont {L.~L.}\ \bibnamefont
			{Gutierrez-Martinez}},\ and\ \bibinfo {author} {\bibfnamefont
			{M.}~\bibnamefont {Sandoval}},\ }\href
	{https://doi.org/10.1088/1742-5468/abfcbb} {\bibfield  {journal} {\bibinfo
			{journal} {J. Stat. Mech.: Theory Exp.}\ }\textbf {\bibinfo {volume}
			{2021}}\bibinfo  {number} { (5)},\ \bibinfo {pages} {053404}}\BibitemShut
	{NoStop}%
	\bibitem [{\citenamefont {Leoni}\ \emph {et~al.}(2020)\citenamefont {Leoni},
		\citenamefont {Paoluzzi}, \citenamefont {Eldeen}, \citenamefont {Estrada},
		\citenamefont {Nguyen}, \citenamefont {Alexandrescu}, \citenamefont {Sherb},\
		and\ \citenamefont {Ahmed}}]{leoni2020surfing}%
	\BibitemOpen
	\bibfield  {number} {  }\bibfield  {author} {\bibinfo {author} {\bibfnamefont
			{M.}~\bibnamefont {Leoni}}, \bibinfo {author} {\bibfnamefont
			{M.}~\bibnamefont {Paoluzzi}}, \bibinfo {author} {\bibfnamefont
			{S.}~\bibnamefont {Eldeen}}, \bibinfo {author} {\bibfnamefont
			{A.}~\bibnamefont {Estrada}}, \bibinfo {author} {\bibfnamefont
			{L.}~\bibnamefont {Nguyen}}, \bibinfo {author} {\bibfnamefont
			{M.}~\bibnamefont {Alexandrescu}}, \bibinfo {author} {\bibfnamefont
			{K.}~\bibnamefont {Sherb}},\ and\ \bibinfo {author} {\bibfnamefont {W.~W.}\
			\bibnamefont {Ahmed}},\ }\href
	{https://doi.org/10.1103/PhysRevResearch.2.043299} {\bibfield  {journal}
		{\bibinfo  {journal} {Phys. Rev. Res.}\ }\textbf {\bibinfo {volume} {2}},\
		\bibinfo {pages} {043299} (\bibinfo {year} {2020})}\BibitemShut {NoStop}%
	\bibitem [{\citenamefont {Horvath}\ \emph {et~al.}(2023)\citenamefont
		{Horvath}, \citenamefont {Slab{\`y}}, \citenamefont {Tomori}, \citenamefont
		{Hovan}, \citenamefont {Miskovsky},\ and\ \citenamefont
		{B{\'a}n{\'o}}}]{horvath2023bouncing}%
	\BibitemOpen
	\bibfield  {author} {\bibinfo {author} {\bibfnamefont {D.}~\bibnamefont
			{Horvath}}, \bibinfo {author} {\bibfnamefont {C.}~\bibnamefont {Slab{\`y}}},
		\bibinfo {author} {\bibfnamefont {Z.}~\bibnamefont {Tomori}}, \bibinfo
		{author} {\bibfnamefont {A.}~\bibnamefont {Hovan}}, \bibinfo {author}
		{\bibfnamefont {P.}~\bibnamefont {Miskovsky}},\ and\ \bibinfo {author}
		{\bibfnamefont {G.}~\bibnamefont {B{\'a}n{\'o}}},\ }\href
	{https://doi.org/10.1103/PhysRevE.107.024603} {\bibfield  {journal} {\bibinfo
			{journal} {Phys. Rev. E}\ }\textbf {\bibinfo {volume} {107}},\ \bibinfo
		{pages} {024603} (\bibinfo {year} {2023})}\BibitemShut {NoStop}%
	\bibitem [{\citenamefont {Chen}\ \emph {et~al.}(2023)\citenamefont {Chen},
		\citenamefont {Welch}, \citenamefont {Leishangthem}, \citenamefont {Ghosh},
		\citenamefont {Zhang}, \citenamefont {Sun}, \citenamefont {Klukas},
		\citenamefont {Tu}, \citenamefont {Cheng},\ and\ \citenamefont
		{Xu}}]{chen2023molecular}%
	\BibitemOpen
	\bibfield  {author} {\bibinfo {author} {\bibfnamefont {L.}~\bibnamefont
			{Chen}}, \bibinfo {author} {\bibfnamefont {K.~J.}\ \bibnamefont {Welch}},
		\bibinfo {author} {\bibfnamefont {P.}~\bibnamefont {Leishangthem}}, \bibinfo
		{author} {\bibfnamefont {D.}~\bibnamefont {Ghosh}}, \bibinfo {author}
		{\bibfnamefont {B.}~\bibnamefont {Zhang}}, \bibinfo {author} {\bibfnamefont
			{T.-P.}\ \bibnamefont {Sun}}, \bibinfo {author} {\bibfnamefont
			{J.}~\bibnamefont {Klukas}}, \bibinfo {author} {\bibfnamefont
			{Z.}~\bibnamefont {Tu}}, \bibinfo {author} {\bibfnamefont {X.}~\bibnamefont
			{Cheng}},\ and\ \bibinfo {author} {\bibfnamefont {X.}~\bibnamefont {Xu}},\
	}\href {https://arxiv.org/abs/2302.10525} {\bibfield  {journal} {\bibinfo
			{journal} {arXiv e-prints}\ } (\bibinfo {year} {2023})}\BibitemShut {NoStop}%
	\bibitem [{\citenamefont {Hamon}\ \emph {et~al.}(2010)\citenamefont {Hamon},
		\citenamefont {Gautier},\ and\ \citenamefont {Garrec}}]{Hamon/etal:2010}%
	\BibitemOpen
	\bibfield  {author} {\bibinfo {author} {\bibfnamefont {P.}~\bibnamefont
			{Hamon}}, \bibinfo {author} {\bibfnamefont {M.}~\bibnamefont {Gautier}},\
		and\ \bibinfo {author} {\bibfnamefont {P.}~\bibnamefont {Garrec}},\ }in\
	\href {https://doi.org/10.1109/IROS.2010.5649189} {\emph {\bibinfo
			{booktitle} {Proceedings of the 2010 IEEE/RSJ International Conference on Intelligent Robots and
				Systems, Taipei, 2010}}}\ (IEEE, New York, \bibinfo {year} {2010})\ pp.\ \bibinfo {pages}
	{6187--6193}\BibitemShut {NoStop}%
	\bibitem [{\citenamefont {Kudrolli}(2010)}]{kudrolli2010concentration}%
	\BibitemOpen
	\bibfield  {author} {\bibinfo {author} {\bibfnamefont {A.}~\bibnamefont
			{Kudrolli}},\ }\href {https://doi.org/10.1103/PhysRevLett.104.088001}
	{\bibfield  {journal} {\bibinfo  {journal} {Phys. Rev. Lett.}\ }\textbf
		{\bibinfo {volume} {104}},\ \bibinfo {pages} {088001} (\bibinfo {year}
		{2010})}\BibitemShut {NoStop}%
	\bibitem [{\citenamefont {Deseigne}\ \emph {et~al.}(2010)\citenamefont
		{Deseigne}, \citenamefont {Dauchot},\ and\ \citenamefont
		{Chat{\'e}}}]{deseigne2010collective}%
	\BibitemOpen
	\bibfield  {author} {\bibinfo {author} {\bibfnamefont {J.}~\bibnamefont
			{Deseigne}}, \bibinfo {author} {\bibfnamefont {O.}~\bibnamefont {Dauchot}},\
		and\ \bibinfo {author} {\bibfnamefont {H.}~\bibnamefont {Chat{\'e}}},\ }\href
	{https://doi.org/10.1103/PhysRevLett.105.098001} {\bibfield  {journal}
		{\bibinfo  {journal} {Phys. Rev. Lett.}\ }\textbf {\bibinfo {volume} {105}},\
		\bibinfo {pages} {098001} (\bibinfo {year} {2010})}\BibitemShut {NoStop}%
	\bibitem [{\citenamefont {Soni}\ \emph {et~al.}(2020)\citenamefont {Soni},
		\citenamefont {Kumar}, \citenamefont {Nambisan}, \citenamefont {Gupta},
		\citenamefont {Sood},\ and\ \citenamefont {Ramaswamy}}]{soni2020phases}%
	\BibitemOpen
	\bibfield  {author} {\bibinfo {author} {\bibfnamefont {H.}~\bibnamefont
			{Soni}}, \bibinfo {author} {\bibfnamefont {N.}~\bibnamefont {Kumar}},
		\bibinfo {author} {\bibfnamefont {J.}~\bibnamefont {Nambisan}}, \bibinfo
		{author} {\bibfnamefont {R.~K.}\ \bibnamefont {Gupta}}, \bibinfo {author}
		{\bibfnamefont {A.}~\bibnamefont {Sood}},\ and\ \bibinfo {author}
		{\bibfnamefont {S.}~\bibnamefont {Ramaswamy}},\ }\href
	{https://doi.org/10.1039/C9SM02552A} {\bibfield  {journal} {\bibinfo
			{journal} {Soft Matter}\ }\textbf {\bibinfo {volume} {16}},\ \bibinfo {pages}
		{7210} (\bibinfo {year} {2020})}\BibitemShut {NoStop}%
	\bibitem [{\citenamefont {Abubakar}\ \emph {et~al.}(2020)\citenamefont
		{Abubakar}, \citenamefont {Yilbas}, \citenamefont {Al-Qahtani}, \citenamefont
		{Alzaydi},\ and\ \citenamefont {Alhelou}}]{Abubakar/etal:2020}%
	\BibitemOpen
	\bibfield  {author} {\bibinfo {author} {\bibfnamefont {A.~A.}\ \bibnamefont
			{Abubakar}}, \bibinfo {author} {\bibfnamefont {B.~S.}\ \bibnamefont
			{Yilbas}}, \bibinfo {author} {\bibfnamefont {H.}~\bibnamefont {Al-Qahtani}},
		\bibinfo {author} {\bibfnamefont {A.}~\bibnamefont {Alzaydi}},\ and\ \bibinfo
		{author} {\bibfnamefont {S.}~\bibnamefont {Alhelou}},\ }\href
	{https://doi.org/10.1038/s41598-020-71356-5} {\bibfield  {journal} {\bibinfo
			{journal} {Sci.~Rep.}\ }\textbf {\bibinfo {volume} {10}},\ \bibinfo
		{pages} {14346} (\bibinfo {year} {2020})}\BibitemShut {NoStop}%
	\bibitem [{\citenamefont {Maslov}\ and\ \citenamefont
		{Fedoriuk}(1981)}]{Maslov1981}%
	\BibitemOpen
	\bibfield  {author} {\bibinfo {author} {\bibfnamefont {V.~P.}\ \bibnamefont
			{Maslov}}\ and\ \bibinfo {author} {\bibfnamefont {M.~V.}\ \bibnamefont
			{Fedoriuk}},\ }\href@noop {} {\emph {\bibinfo {title} {Semiclassical
				Approximation in Quantum Mechanics}}}\ (\bibinfo  {publisher} {Reidel},\
	\bibinfo {address} {Dordrecht},\ \bibinfo {year} {1981})\BibitemShut
	{NoStop}%
	\bibitem [{\citenamefont {Antonov}\ \emph {et~al.}(2023)\citenamefont
		{Antonov}, \citenamefont {Leonidov},\ and\ \citenamefont
		{Semenov}}]{Antonov2023}%
	\BibitemOpen
	\bibfield  {author} {\bibinfo {author} {\bibfnamefont {A.}~\bibnamefont
			{Antonov}}, \bibinfo {author} {\bibfnamefont {A.}~\bibnamefont {Leonidov}},\
		and\ \bibinfo {author} {\bibfnamefont {A.}~\bibnamefont {Semenov}},\ }\href
	{https://doi.org/10.1103/PhysRevE.108.024134} {\bibfield  {journal} {\bibinfo
			{journal} {Phys. Rev. E}\ }\textbf {\bibinfo {volume} {108}},\ \bibinfo
		{pages} {024134} (\bibinfo {year} {2023})}\BibitemShut {NoStop}%
	\bibitem [{\citenamefont {Landau}\ and\ \citenamefont
		{Lifshitz}(1976)}]{Landau}%
	\BibitemOpen
	\bibfield  {author} {\bibinfo {author} {\bibfnamefont {L.~D.}\ \bibnamefont
			{Landau}}\ and\ \bibinfo {author} {\bibfnamefont {E.~M.}\ \bibnamefont
			{Lifshitz}},\ }\href@noop {} {\emph {\bibinfo {title} {Mechanics. Vol. 1}}}\
	(\bibinfo  {publisher} {Butterworth-Heinemann},\ \bibinfo {year}
	{1976})\BibitemShut {NoStop}%
\end{thebibliography}

\begin{thebibliography}{5}%
	\makeatletter
	\providecommand \@ifxundefined [1]{%
		\@ifx{#1\undefined}
	}%
	\providecommand \@ifnum [1]{%
		\ifnum #1\expandafter \@firstoftwo
		\else \expandafter \@secondoftwo
		\fi
	}%
	\providecommand \@ifx [1]{%
		\ifx #1\expandafter \@firstoftwo
		\else \expandafter \@secondoftwo
		\fi
	}%
	\providecommand \natexlab [1]{#1}%
	\providecommand \enquote  [1]{``#1''}%
	\providecommand \bibnamefont  [1]{#1}%
	\providecommand \bibfnamefont [1]{#1}%
	\providecommand \citenamefont [1]{#1}%
	\providecommand \href@noop [0]{\@secondoftwo}%
	\providecommand \href [0]{\begingroup \@sanitize@url \@href}%
	\providecommand \@href[1]{\@@startlink{#1}\@@href}%
	\providecommand \@@href[1]{\endgroup#1\@@endlink}%
	\providecommand \@sanitize@url [0]{\catcode `\\12\catcode `\$12\catcode
		`\&12\catcode `\#12\catcode `\^12\catcode `\_12\catcode `\%12\relax}%
	\providecommand \@@startlink[1]{}%
	\providecommand \@@endlink[0]{}%
	\providecommand \url  [0]{\begingroup\@sanitize@url \@url }%
	\providecommand \@url [1]{\endgroup\@href {#1}{\urlprefix }}%
	\providecommand \urlprefix  [0]{URL }%
	\providecommand \Eprint [0]{\href }%
	\providecommand \doibase [0]{https://doi.org/}%
	\providecommand \selectlanguage [0]{\@gobble}%
	\providecommand \bibinfo  [0]{\@secondoftwo}%
	\providecommand \bibfield  [0]{\@secondoftwo}%
	\providecommand \translation [1]{[#1]}%
	\providecommand \BibitemOpen [0]{}%
	\providecommand \bibitemStop [0]{}%
	\providecommand \bibitemNoStop [0]{.\EOS\space}%
	\providecommand \EOS [0]{\spacefactor3000\relax}%
	\providecommand \BibitemShut  [1]{\csname bibitem#1\endcsname}%
	\let\auto@bib@innerbib\@empty
	\bibitem [{\citenamefont {Scholz}\ \emph {et~al.}(2018)\citenamefont {Scholz},
		\citenamefont {Jahanshahi}, \citenamefont {Ldov},\ and\ \citenamefont
		{L{\"o}wen}}]{scholz2018inertialS}%
	\BibitemOpen
	\bibfield  {author} {\bibinfo {author} {\bibfnamefont {C.}~\bibnamefont
			{Scholz}}, \bibinfo {author} {\bibfnamefont {S.}~\bibnamefont {Jahanshahi}},
		\bibinfo {author} {\bibfnamefont {A.}~\bibnamefont {Ldov}},\ and\ \bibinfo
		{author} {\bibfnamefont {H.}~\bibnamefont {L{\"o}wen}},\ }\href
	{https://doi.org/10.1038/s41467-018-07596-x} {\bibfield  {journal} {\bibinfo
			{journal} {Nat. Commun.}\ }\textbf {\bibinfo {volume} {9}},\ \bibinfo {pages}
		{5156} (\bibinfo {year} {2018})}\BibitemShut {NoStop}%
	\bibitem [{\citenamefont {Caroli}\ \emph {et~al.}(1981)\citenamefont {Caroli},
		\citenamefont {Caroli},\ and\ \citenamefont {Roulet}}]{Caroli1981S}%
	\BibitemOpen
	\bibfield  {author} {\bibinfo {author} {\bibfnamefont {B.}~\bibnamefont
			{Caroli}}, \bibinfo {author} {\bibfnamefont {C.}~\bibnamefont {Caroli}},\
		and\ \bibinfo {author} {\bibfnamefont {B.}~\bibnamefont {Roulet}},\ }\href
	{https://doi.org/10.1007/BF01106788} {\bibfield  {journal} {\bibinfo
			{journal} {J. Stat. Phys.}\ }\textbf {\bibinfo {volume} {26}},\ \bibinfo
		{pages} {83} (\bibinfo {year} {1981})}\BibitemShut {NoStop}%
	\bibitem [{\citenamefont {Maslov}\ and\ \citenamefont
		{Fedoriuk}(1981)}]{Maslov1981S}%
	\BibitemOpen
	\bibfield  {author} {\bibinfo {author} {\bibfnamefont {V.~P.}\ \bibnamefont
			{Maslov}}\ and\ \bibinfo {author} {\bibfnamefont {M.~V.}\ \bibnamefont
			{Fedoriuk}},\ }\href@noop {} {\emph {\bibinfo {title} {Semiclassical
				Approximation in Quantum Mechanics}}}\ (\bibinfo  {publisher} {Reidel},\
	\bibinfo {address} {Dordrecht},\ \bibinfo {year} {1981})\BibitemShut
	{NoStop}%
	\bibitem [{\citenamefont {Antonov}\ \emph {et~al.}(2023)\citenamefont
		{Antonov}, \citenamefont {Leonidov},\ and\ \citenamefont
		{Semenov}}]{Antonov2023S}%
	\BibitemOpen
	\bibfield  {author} {\bibinfo {author} {\bibfnamefont {A.}~\bibnamefont
			{Antonov}}, \bibinfo {author} {\bibfnamefont {A.}~\bibnamefont {Leonidov}},\
		and\ \bibinfo {author} {\bibfnamefont {A.}~\bibnamefont {Semenov}},\ }\href
	{https://doi.org/10.1103/PhysRevE.108.024134} {\bibfield  {journal} {\bibinfo
			{journal} {Phys. Rev. E}\ }\textbf {\bibinfo {volume} {108}},\ \bibinfo
		{pages} {024134} (\bibinfo {year} {2023})}\BibitemShut {NoStop}%
	\bibitem [{\citenamefont {Landau}\ and\ \citenamefont
		{Lifshitz}(1976)}]{LandauS}%
	\BibitemOpen
	\bibfield  {author} {\bibinfo {author} {\bibfnamefont {L.~D.}\ \bibnamefont
			{Landau}}\ and\ \bibinfo {author} {\bibfnamefont {E.~M.}\ \bibnamefont
			{Lifshitz}},\ }\href@noop {} {\emph {\bibinfo {title} {Mechanics. Vol. 1}}}\
	(\bibinfo  {publisher} {Butterworth-Heinemann},\ \bibinfo {year}
	{1976})\BibitemShut {NoStop}%
\end{thebibliography}
\end{document}